\documentclass[preprint2]{aastex}

\shorttitle{Stellar Wobble by a Planet}
\shortauthors{Takeuchi, Velusamy, \& Lin}

\begin{document}


\title{Apparent Stellar Wobble by a Planet in a Circumstellar Disk:
Limitations on Planet Detection by Astrometry \footnote{Accepted by ApJ}}

\author{Taku Takeuchi\altaffilmark{2,3}, T. Velusamy\altaffilmark{4}, and
  D. N. C. Lin\altaffilmark{3}} 
\altaffiltext{2}{Earth and Planetary Sciences, Kobe University, Kobe
  657-8501, Japan ; taku@kobe-u.ac.jp}
\altaffiltext{3}{UCO/Lick Observatory, University of California, Santa Cruz,
CA95064; lin@ucolick.org}
\altaffiltext{4}{Jet Propulsion Laboratory, California Institute of
  Technology, MS 169-506, Pasadena, CA91109}

\begin{abstract}

Astrometric detection of a stellar wobble on the plane of the sky will
provide us a next breakthrough in searching extrasolar planets.
The Space Interferometry Mission (SIM) is expected to achieve a
high-precision astrometry as accurate as 1 $\mu$as, which is
precise enough to discover a new-born Jupiter mass planet around a
pre-main-sequence (PMS) star in the Taurus-Auriga star forming region.
PMS stars, however, have circum-stellar disks that may be obstacles to
the precise measurement of the stellar position.
We present results on disk influences to the stellar wobble.
The density waves excited by a planet move both of the disk's mass
center and the photo-center.
The motion of the disk mass center induces an additional wobble of the
stellar position, and the motion of the disk photo-center causes a
contamination in the measurement of the stellar position.
We show that the additional stellar motion dynamically caused by the
disk's gravity is always negligible, but that the contamination of the
disk light can interfere with the precise measurement of the stellar
position, if the planet's mass is smaller than $\sim 10$ Jupiter mass.
The motion of the disk photo-center is sensitive to a slight
change in the wave pattern and the disk properties.
Measurements by interferometers are generally insensitive to extended
sources such as disks. Because of this property SIM will not suffer
significant contaminations of the disk light, even if the planet's
mass is as small as 1 Jupiter mass. 

\end{abstract}



\section{Introduction}

Since the first discovery by Mayor \& Queloz (1995), more than 120
extrasolar planets around main-sequence stars have been counted in the
list (Schneider 2004{\footnote{http://www.obspm.fr/encycl/encycl.html}).
Most of them were discovered by radial velocity measurement of
stars (see, e.g., Marcy \& Butler 1998; Perryman 2000).
One of the planets that were already found by the radial velocity
measurement has been confirmed to have transits on the star (Charbonneau et
al. 2000), and a few planets were originally discovered from their
transits (Konacki et al. 2003a,b).
A gravitational microlensing event was reported that it had been caused
by a planet orbiting around binary stars (Bennett et al. 1999).

It is considered that astrometric detection of wobbling stars will be a
strong method for discovering planets in the near future.
A star orbiting the center of mass of the star-planet system draws an
ellipse on the plane of the sky (after the parallax and the stellar
proper motion are subtracted).
The semi-major axis (in arcsec) of the ellipse caused by a planet having
a circular orbit is 
\begin{equation}
a = \frac{M_p}{M+M_p} \frac{r_p/{\rm AU}}{d/{\rm pc}} \ ,
\end{equation}
where $M_p$ and $M$ are the masses of the planet and the star,
respectively, $r_p$ is the distance between the planet and the star, and
$d$ is the distance of the star from the observer.
When a solar mass star with a Jupiter mass planet orbiting at 5 AU is
observed from 10 pc, the amplitude of the stellar wobble is about 500
$\mu$as.

Measurement of stellar positions with a relative accuracy of
sub-milliarcsec is a challenge for ground-based optical observations.
Recent improvement of interferometers has made it possible, and
planet survey projects are ongoing at the Keck Interferometer and the
VLTI interferometer (for reviews, Perryman 2000; Monnier 2003).

The Space Interferometry Mission (SIM), which is scheduled for launch
in 2009, is expected to make a significant progress of high-precision
astrometry (Danner \& Unwin 1999).  
The SIM is expected to achieve $\sim 1 \ \mu$as relative accuracy with its
narrow-angle astrometric mode.
After the 5 yr mission duration, it can detect a Jupiter mass
planet at 300 pc, or a Neptune mass planet at 20 pc, if it orbits at
1-5 AU around a 1 $M_{\sun}$ star (Sozzetti et al. 2002).
This high accuracy provides us an opportunity of searching new-born
planets in star forming regions.
For example, the Taurus-Auriga molecular cloud, which is at 140 pc and has
more than 200 T Tauri stars (Kenyon \& Hartmann 1995), will be a good
site for searching a young planet.
Discovery of a young planet around a pre-main-sequence (PMS) star will
tell us when planet formation occurs.
The standard scenario of planet formation assumes that planets form
through coagulations of planetesimals (Hayashi, Nakazawa, \& Nakagawa
1985; Lissauer 1993).
According to an early version of this model (Pollack et al. 1996), it
may take more than $10^7$ yr to form a Jupiter mass planet at 5 AU.
However, recent results on the oligarchic growth of planet cores
(Kokubo \& Ida 1998, 2000; Thommes, Duncan, \& Levison 2003) and
considerations of dust opacity in a protoplanet's envelope (Ikoma,
Nakazawa, \& Emori 2000; Podolak 2003) indicate that the formation of
Jupiter may take much less time. 
In relatively massive disks where the asymptotic
core masses are large, gas giant planets can also form rapidly and
readily.  But in these disks, gravitational instability may also lead to
the rapid formation of gas giant (Boss 1997).  Although the formation
time scale cannot be used to directly distinguish the paths of planet
formation, dynamical diversity may provide useful clues.  In the
gravitational instability scenario, planets are expected to form with
large masses and eccentricities.  In contrast, core accretion is
expected to proceed in a tranquil environment with modest mass and low
eccentricity.  In the later scenario, planets may acquire their dynamical
diversity during the disk evolution (Ida \& Lin 2004) and depletion
(Nagasawa, Lin, \& Ida 2003).  
A systematic comparison between dynamical
properties of protoplanets around PMS stars and those of mature planets
around main sequence stars provides a useful diagnostic on both the
formation and evolution of planets and planetary systems.
For example, Boss (1998) and Rice et al. (2003) calculated stellar
wobbles induced by gravitationally unstable disks.
Their models showed that the disk fragmentation induces significantly
larger or faster wobbles than the disks without fragmentation.

PMS stars, however, generally have circum-stellar disks.
The disk may be a obstacle to the planet search using astrometry.
Firstly, the planet disturbs the disk as well as the star, and moves the
position of the disk center off the star.
The gravity of the disturbed disk induces an extra motion of the star, in
addition to the direct wobbling by the planet.
Secondly, the disturbance of the disk breaks the symmetry of the disk
emission with regard to the star.
The spiral waves excited by the planet make a wavy shape of
the disk surface, leading to lighting and shadowing the surface.
The scattered light image of the disk is disturbed so as to have a
spiral pattern, and the center of the disk light is moved away from the star.
When the position of the star is measured by a small telescope (for
example, the telescopes of the SIM are 33 cm diameter), the beam size is
much larger than the angular diameter of the disk, thus the
observable value is only the central position of the light from the
star-disk system.
If the shift of the photo-center of the system is similar or larger than
the amplitude of the dynamical stellar wobble, it is difficult to
extract the dynamical motion from the apparent motion of the photo-center.

In order to examine how a circum-stellar disk affects the precise
measurement of the stellar position, we study the stellar wobble induced
by a disk both dynamically (through its mass distribution) and
illuminationally (through its scattered light distribution).
In \S2, we calculate linear responses of the disk to the
planet's gravity to obtain a spiral wave pattern.
For simplicity, the disk is treated as two-dimensional and the planet's
orbit is circular.
In \S3, we examine the dynamical effect of the disk.
The gravity exerted on the star by the disturbed disk is calculated, and
the stellar wobble induced by the disk is compared to that by the planet. 
We show that the disk's dynamical effect can be neglected.
In \S4, the motion of the central position of disk light is calculated.
We assume that the three-dimensional wavy pattern of the disk surface
relates to the surface density fluctuation calculated in \S2, where the
disk is treated as two-dimensional.
The lighting and shadowing the disk surface by starlight are
calculated to make an image of the scattered light.
The shift of the photo-center is calculated and compared to
the dynamical motion of the star.
In \S5, we apply our result to the SIM. 
We simulate observations by the SIM and calculate the anticipated
astrometry shift induced by the disk light.


\section{Density Wave Excitation by a Planet}

\subsection{Wave Equations}

\subsubsection{Unperturbed Disk}

We consider a system composed of a central star, a circumstellar disk,
and a planet in the disk.
The planet's orbit is assumed to be circular with an angular velocity
\begin{equation}
\Omega_p = \left( \frac{G M}{r_p^3} \right)^{1/2} \ ,
\end{equation}
where we ignore contributions from the
planet's mass $M_p$ and the disk mass $M_d$, assuming $M_p$, $M_d \ll M$.

A planet embedded in a gas disk opens a gap around its orbit, provided
that the planet's mass is larger than the Neptune or Saturn's mass [$ (1-3)
  \times 10^{-4} M_{\sun}$] (Takeuchi, Miyama, \& Lin 1996).
The gap width is in general a function of the planet's mass, the gas
sound speed, and the gas viscosity.
For a small mass planet like Neptune, the gap width is of order of the
disk thickness, while for a Jupiter mass planet, the gap spreads from
the $m=2$ inner Lindblad resonance (at $0.63 r_p$) to the $m=2$ outer
Lindblad resonance (at $1.31 r_p$; Lin \& Papaloizou 1986; Takeuchi et
al. 1996).
In this work, we consider an unperturbed surface density distribution of
a disk having a gap, and for simplicity, it is written as
\begin{equation}
\sigma_0 = f_{\rm gap} \sigma_* r_{\rm AU}^{-1} \ ,
\label{eq:sigma0}
\end{equation}
where $r_{\rm AU}$ is the non-dimensional radial coordinate normalized by
1 AU, and $\sigma_*$ is the surface density at 1 AU.
We assume the surface density decreases with $r$ as a power law with an
index 1, and near the planet's orbit it drops; $f_{\rm gap}$
is unity except for the planet's neighborhood at which it drops to zero.
We use an approximate expression for the gap shape:
\begin{equation}
f_{\rm gap} = \frac{\exp(\pm 2x)}{\exp(\pm 2x)+1} \ ,
\label{eq:gapshape}
\end{equation}
where the upper and lower signs are for the disk outside the
planet's orbit and for the inner disk, respectively, and the
non-dimensional distance $x$ from the gap edge is 
\begin{equation}
x = \frac{r - r_{g,\rm out}}{\Delta_{\rm edge}} ,
\end{equation}
for the outer disk and 
\begin{equation}
x = \frac{r - r_{g,\rm in}}{\Delta_{\rm edge}} ,
\end{equation}
for the inner disk.
The median locations of the gap edges, $r_{g,\rm out}$ and $r_{g,\rm
in}$, are expediently defined as the points where the surface density
drops to half the value of the no-gap disk.
The edge width $\Delta_{\rm edge}$ is of order of the disk half-thickness $h$.
The half-width of the gap is $\Delta_g = (r_{g, \rm out}-r_{g,\rm in})/2$.
We assume that the gap shape is symmetrical with regard to the planet's
position and that $\Delta_g = r_{g, \rm out}-r_p = r_p - r_{g, \rm in}$.

A polytropic equation of state with an index $\gamma$ is assumed for the
gas.
The two-dimensional pressure is $P_0 = K \sigma_0^{\gamma}$.
The enthalpy and the sound speed are written by 
\begin{equation}
\eta_0 = \frac{\gamma}{\gamma-1} K \sigma_0^{\gamma-1} \ ,
\end{equation}
\begin{equation}
c_0^2 = (\gamma-1) \eta_0 \ ,
\end{equation}
respectively.
The gas rotation law is assumed as Keplerian: $\Omega^2 = GM/r^3$.
Though in actual disks the gas pressure makes the rotation slightly
slower than Keplerian, we neglect this small difference.
For a given enthalpy distribution, the half-thickness of the unperturbed
disk is $h=\sqrt{2 \eta_0}/\Omega$.
The derivation of this equation is discussed in \S\ref{sec:thickness} below.

\subsubsection{Perturbation by a Planet}

We solve the gas motions perturbed by the planet's gravity.
The gap around the planet's orbit divides the disk into two parts.
The perturbation equations are solved separately for the inner and outer
disks.
We use cylindrical coordinates $(r,\theta)$.
For the inner disk the coordinate origin is set on the
central star, and the coordinates are not an inertial frame.
On the other hand, for the outer disk the origin is on the center of
mass of the system.
We neglect the self-gravity of the disk.

The planet's perturbation potential on the inner disk is written as
\begin{eqnarray}
\phi_1 & = & -\frac{G M_p}{(r_p^2 + r^2 - 2 r_p r \cos
(\theta-\Omega_p t))^{1/2}} \nonumber \\
 & + & \frac{G M_p r \cos (\theta-\Omega_p t)}{r_p^2} \ .
\end{eqnarray}
The second term is the indirect term that represents the inertial force.
For the outer disk, the star also orbits around the origin and causes a
perturbation.
The perturbation potential is 
\begin{eqnarray}
\phi_1 & = & -\frac{G M}{[r_a^2 + r^2 - 2 r_a r \cos
(\theta-\Omega_p t - \pi)]^{1/2}} \nonumber \\
 & - & \frac{G M_p}{[r_b^2 + r^2 - 2 r_b r \cos (\theta-\Omega_p t)]^{1/2}} 
\ ,
\end{eqnarray}
where $r_a=r_p M_p/(M+M_p)$ and $r_b=r_p M/(M+M_p)$ are the distances
from the origin to the star and to the planet, respectively.

We solve for Fourier components of the perturbations. 
Each perturbation variable, which is with subscript ``1'', is decomposed
in the azimuthal direction as 
\begin{equation}
X_1 (r,\theta) = \sum_{m=0}^{\infty} {\rm Re} [ X_{1,m} (r) \exp [ im (\theta -
\Omega_p t) ] \ ,
\label{eq:Fourier}
\end{equation}
where $m$ is the wavenumber in the $\theta$-direction.
In the following part of the paper, we solve Fourier components
$X_{1,m}$, but we omit subscript ``$m$'' in writing whenever this
omission does not cause any confusion.

The perturbation equations are as follows (Goldreich \& Tremaine 1979):
\begin{equation}
i m (\Omega - \Omega_p) u_1 - 2 \Omega v_1 = -
\frac{d}{d r} ( \phi_1 + \eta_1 ) \ ,
\label{eq:perturb_1}
\end{equation}
\begin{equation}
i m (\Omega - \Omega_p) v_1 + 2 B u_1 = - \frac{i m}{r} ( \phi_1 +
\eta_1 ) \ , 
\label{eq:perturb_2}
\end{equation}
\begin{equation}
i m (\Omega - \Omega_p) \sigma_1 + \frac{1}{r} \frac{d}{d
r} ( r \sigma_0 u_1) + \frac{i m \sigma_0}{r} v_1 = 0 \ ,
\label{eq:perturb_3}
\end{equation}
where $u_1$ and $v_1$ are respectively the radial and azimuthal
components of the velocity perturbation, and Oort's $B=\Omega + (r/2) (d
\Omega / dr)$.
The enthalpy perturbation relates to $\sigma_1$ as
\begin{equation}
\eta_1 = c_0^2 \frac{\sigma_1}{\sigma_0} \ .
\end{equation}

The Fourier transformed perturbation potential $\phi_1$ for the inner
disk is expressed as
\begin{equation}
\phi_1 (r) = 
- \frac{G M_p (2-\delta_{m0})}{2 r_p} b_{1/2}^m (\beta) +  \frac{G
  M_p r}{r_p^2} \delta_{m1} \ ,
\label{eq:phi1in}
\end{equation}
where $\beta=r/r_p$, $\delta_{mn}$ is the Kronecker's delta, and $
b_{1/2}^m$ is the Laplace coefficient,
\begin{equation}
b_{1/2}^m (\beta) = \frac{2}{\pi} \int_0^{\pi} \frac{\cos m \phi d \phi}
{(1-2 \beta \cos \phi + \beta^2)^{1/2}} \ .
\label{eq:laplace}
\end{equation}
For the outer disk, 
\begin{eqnarray}
\phi_1 (r) & = & 
(-)^{m-1} \frac{G M (2-\delta_{m0})}{2 r_a} b_{1/2}^m (\beta_a)
\nonumber \\
& -& \frac{G M_p (2-\delta_{m0})}{2 r_b} b_{1/2}^m (\beta_b) \ ,
\label{eq:phi1out}
\end{eqnarray}
where $\beta_a=r/r_a$ and $\beta_b=r/r_b$.

\subsection{Numerical Integration}

We expect that the planet excites steady waves in the disk, i.e., in the
frame rotating with the planet, the wave pattern keeps the same
shape.
In such a steady state, energy input to the waves from the planet is
balanced by energy loss of wave dissipation caused by shocks or the 
viscous damping.
This situation is numerically expressed by allowing the waves to freely
escape from the boundaries of the computational domain, i.e., we take
the radiative boundary condition.

We solve the perturbation equations of the inner and outer disks separately.
In solving the inner disk, the numerical integration starts from the
point $r_{\rm in,2}$ near the planet, at which the gas density is
sufficiently low.
Equations (\ref{eq:perturb_1})-(\ref{eq:perturb_3}) are
solved inward to a certain radius $r_{\rm in,1}$, where the waves
wind tightly enough.
For the outer disk, the computation is performed for $[r_{\rm out,1},
r_{\rm out, 2}]$.

The computational boundaries of planet's side are at the gap edges,
$r_{\rm in,2} = r_{g, \rm in} + 2 \Delta_{\rm edge}$ for the inner disk,
and $r_{\rm out,1} = 
r_{g, \rm out} - 2 \Delta_{\rm edge}$ for the outer disk.
The densities at these boundaries are about 2\% of the non-gap values
(eq. [\ref{eq:gapshape}]).
We adopt the free boundary condition.
The Lagrange derivative of the enthalpy is zero: $D \eta / Dt = 0$.
Assuming that the radial derivative of the unperturbed enthalpy is zero, we
simply have $\eta_1 = 0$.

Equations (\ref{eq:perturb_1})-(\ref{eq:perturb_3}) are integrated using
the fourth-order Runge-Kutta method.
The Laplace coefficients (\ref{eq:laplace}) are numerically calculated
using the method by Korycansky \& Pollack (1993).
The boundary condition at $r_{\rm in,1}$ or $r_{\rm out,2}$ is 
radiative.
At these points, only outgoing wave is allowed to exist.
In a WKB approximation, the enthalpy perturbation is expressed as
$\eta_1 = H(r) \exp [ \int i k dr]$, where $k=\pm \sqrt{-D}/c_0$ and
$D=\Omega^2 - m^2 (\Omega-\Omega_p)^2$ (eq. [19] in Goldreich \&
Tremaine (1979) with $G=0$, $\omega=m \Omega_p$, and $\kappa=\Omega$).
The radiative boundary condition is, as in Korycansky \& Pollack (1993),
\begin{equation}
\frac{d \eta_1}{dr} = \left( \frac{1}{H} \frac{dH}{dr} + ik \right)
\eta_1 \ ,
\end{equation}
where we take a positive sign of the wavenumber $k$ (the negative sign
corresponds to the ingoing wave).
The derivative of the amplitude of $\eta_1$ is calculated by a similar
way deriving equation (18b) in Goldreich \& Tremaine (1979), and is
\begin{equation}
\frac{1}{H} \frac{dH}{dr} = - \frac{1}{2} \frac{d}{dr} \ln \left( \frac{r
\sigma_0 k}{D} \right) \ .
\end{equation}
For the $m=1$ mode in the inner disk, there is no wave-zone (i.e., $k$
is pure imaginary).
We use the free boundary condition ($\eta_1=0$) both at $r_{\rm in,1}$
and $r_{\rm in,2}$.

\subsection{Viscous Damping of Waves}

The calculations in the previous subsections do not take any wave
damping into account.
In actual disks, waves attenuate through viscous dissipation or shock
dissipation.
We consider here viscous wave damping, which works even though the wave
amplitude is not large enough to cause shocks.
Waves are excited near the planet and propagate away. 
As they propagate further, their amplitudes decline through the viscous
dissipation.
We write the enthalpy perturbation in a viscous disk as
\begin{equation}
\eta_{1, \rm vis} = f_{\rm vis} \eta_1 \ ,
\end{equation}
where $\eta_1$ is the enthalpy solved for the non-viscous disk in the
previous subsection, and $f_{\rm vis}$, a factor of viscous damping, is
unity near the planet and goes to zero with the distance from the
planet.
From equations (A10) and (A15) in Takeuchi et al. (1996), the damping factor is
written as
\begin{eqnarray}
f_{\rm vis} & = & \exp \left[ - \int_{r_{L,m}}^{r} \left\{ \zeta + \left(
\frac{4}{3} + \frac{\Omega^2}{m^2 (\Omega - \Omega_p)^2} \right) \nu
\right\} \right. \nonumber \\
& \times & \left.
\frac{m (\Omega_p - \Omega)}{2 c_0^2} k d \tilde{r} \right] ,
\label{eq:fvis}
\end{eqnarray}
where $r_{L,m}$ is the radius of the Lindblad resonance of the $m$-th
mode wave,
and $\zeta$ and $\nu$ are the bulk and shear viscosities, respectively.
For simplicity, we use the $\alpha$ prescription for viscosity: $\zeta = \nu
= \alpha_{\rm vis} c_0^2 / \Omega$. 
The radial wavenumber is $k=\sqrt{-D} / c_0$ for outgoing waves.

\subsection{Wave Properties}

In this subsection, we present properties of the density waves excited
by a planet.
The wave patterns are different from the ones derived by Korycansky \&
Pollack (1993) because of the assumptions that the disk has a gap and
that the waves viscously dissipate.

As a fiducial model, we adopt the following parameters:
The central star's mass is $M=1 M_{\sun}$. 
The planet's orbital radius is $r_p=5$AU.
The surface density profile is $\sigma_0 = 1000 f_{\rm gap} r_{\rm
AU}^{-1} \ {\rm 
g \ cm}^{-3}$ with a gap of the half width $\Delta_g = 0.4 r_p$ and the
edge width $\Delta_{\rm edge} = 5 \times 10^{-2} r_p$.
The disk inner radius is $r_{\rm in,1}=0.01r_p$ and the outer radius is
$r_{\rm out,2}=5r_p$.
The polytrope index is $\gamma = 1.4$.
The sound speed at 1 AU is $c_0(1 \ {\rm AU}) = 10^5 {\rm cm \ s}^{-1}$.
The viscosity parameter is $\alpha_{\rm vis} = 10^{-2}$.
The planet's mass is arbitrary in linear calculations.
We take a Jupiter mass planet ($M_p = 10^{-3} M_{\sun}$) as a typical
value in the following discussions.

Figure \ref{fig:flux} shows which modes are effectively excited.
The angular momentum flux of waves is a convenient value to compare,
and its expression is written by Goldreich \& Tremaine (1979) as
\begin{eqnarray}
F_A &=& \frac{\pi m r \sigma_0}{\Omega^2 - m^2 (\Omega-\Omega_p)^2}
\left[ {\rm Im} (\eta_1) {\rm Re} \frac{d}{dr} (\phi_1+\eta_1) \right.
\nonumber \\
& - & \left.
 {\rm Re} (\phi_1 + \eta_1) {\rm Im} \frac{d \eta_1}{dr} \right] \ ,
\end{eqnarray}
where Re and Im mean the real and imaginary parts, respectively.
Fluxes $F_A$ are evaluated for waves without viscous damping.
For such waves, $F_A$ converges to a certain value at locations far enough from
the planet (see Fig. 7 in Korycansky \& Pollack 1993).
In Figure \ref{fig:flux}, converged values of $F_A$ are plotted.
Except for small fluctuations, the angular momentum flux is a decreasing
function of mode $m$, and waves of $m \la 3$ dominate.
High $m \ga 3$ waves are not excited strongly, because their
Lindblad resonances (LRs) are buried in the gap (Artymowicz \& Lubow
1994).
The locations of LRs are closer to the planet for higher $m$ (for a
schematic illustration, see Fig.1 in Takeuchi et al. 1996).
The resonance positions of the $m=3$ wave are $r_{L,3} = 0.76 r_p$ and $1.21
r_p$, which are inside the gap of half width $\Delta_g = 0.4 r_p$, while
$m=2$ LRs at $0.63 r_p$ and at $1.31 r_p$ are marginally on the gap edges.

Waves cannot freely travel through the disk.
If the disk viscosity is large enough, waves attenuate before they reach the
disk inner or outer boundaries.
Figure \ref{fig:damp} shows the damping factors, $f_{\rm vis}$, for
various $m$ of waves.
Higher $m$ waves dissipate more quickly and lower $m$ waves can travel
further away from the planet.
If the disk viscosity is large ($\alpha_{\rm vis} = 10^{-2}$), however, even
$m=2$ wave cannot reach the central star nor the outer part further than
$4 r_p$.
An $m=2$ wave can reach the central star if the viscosity is as small
as $\alpha_{\rm vis}=10^{-3}$, although it cannot travel further than $8 r_p$.
Even in such a low viscous disk, $m \ga 5$ waves dissipate before
they reach the central star.

The above results are summarized as follows. 
After the opening of a gap by a planet, excitation of high $m$ waves are
suppressed. Only $m \la 3$ waves can be excited. 
These waves may propagate close to the central
star, although they dissipate before arriving at $\sim 4 r_p$ in a viscous
disk of $\alpha_{\rm vis} =10^{-2}$.
The wave propagation changes the disk shape inside $\sim 4 r_p$.
Figure \ref{fig:img_sden} shows the surface density 
perturbation.
The perturbations of $m=1$ to 20 are summed up to construct the wave
pattern.
In the inner disk, however, the $m=2$ wave dominates the wave pattern,
while in the outer disk the $m=1$ wave dominates.
This is because there is no inner LR for $m=1$, and because $m \ga 3$
LRs are buried in the gap.
The wave amplitude declines under the viscous damping at larger
distances from the planet.


\section{Dynamical Stellar Wobble}

A planet wobbles its central star.
While the planet's gravity shifts the stellar position directly, 
if there is a circumstellar disk, the planet affects the stellar
position through altering the disk density distribution as well.
Since the $m=1$ wave pattern does not have any symmetry that
conserves the center of mass of the disk, the disk's mean position moves.
Because the mass center of a system composed of a star, a planet, and
a disk must be conserved, the star moves so as to compensate the motions of
the planet and the disk.

The disk's gravity to the star is considered in this section, although
the disk self-gravity is ignored in the calculation of wave
excitation.
This treatment is justified if the perturbation of the disk's gravitational
potential is much smaller than the gas enthalpy perturbation.
When the Toomre's $Q$ value, which is $c_0 \Omega / (\pi G \sigma_0)$, is
much larger than 1, this condition is satisfied [see also the discussion
by Ward (1986)].
In our model, $Q=19$ at 10 AU.

In the previous sections, different coordinate systems have
been used  for the inner and outer disks.
We use hereafter the coordinates $(r,\theta,z)$ and the corresponding
Cartesian coordinates $(x,y,z)$, whose origin is set to the star.
When we need to refer to coordinates with the origin on the center of
mass, the coordinate variables are distinguished by writing
$(r^{\prime}, \theta^{\prime}, z)$.

The $x$-position of the center of mass of the inner disk is given by
\begin{eqnarray}
x_{d,\rm in} & = & \frac{1}{M_{d, \rm in}} \int_{r_{\rm in,1}}^{r_{\rm in,2}}
 \int_0^{2 \pi} 
 \sum_m {\rm Re} [ \sigma_{1,m} (r) \nonumber \\
& \times &
\exp ( i m (\theta - \Omega_p t) ) ]
 r^2 \cos \theta  d \theta dr \nonumber \\ 
& = & \frac{\pi}{M_{d, \rm in}} \int_{r_{\rm in,1}}^{r_{\rm in,2}} r^2 
[ {\rm Re} (\sigma_{1,1}) \cos ( \Omega_p t) \nonumber \\
& + & {\rm Im} (\sigma_{1,1})
 \sin ( \Omega_p t) ] dr
\ ,
\end{eqnarray}
where $M_{d, \rm in}$ is the inner disk mass, and the subscript ``$m$''
is written explicitly.
Because of the symmetry, all the contributions vanish after
$\theta$-integration except for the $m=1$ mode.
For the outer disk the disk variables are calculated in the inertial
frame, and then
\begin{eqnarray}
x_{d,\rm out} &= &
\frac{\pi}{M_{d, \rm out}} \int_{r_{\rm out,1}^{\prime}}^{r_{\rm
out,2}^{\prime}} r^{\prime 2}
[ {\rm Re} (\sigma_{1,1}) \cos ( \Omega_p t) \nonumber \\
& + & {\rm Im} (\sigma_{1,1})
 \sin ( \Omega_p t) ] dr^{\prime}
+ r_a \cos (\Omega_p t) \ , \ \ \ \ \
\end{eqnarray}
where the second term represents the motion of the inertial coordinates
$(r^{\prime}, \theta^{\prime}, z)$ to the star.
The $y$-positions of the centers of mass are similarly
\begin{eqnarray}
y_{d, \rm in} &=& \frac{\pi}{M_{d, \rm in}}  \int_{r_{\rm in,1}}^{r_{\rm
in,2}} r^2  [ {\rm Re} (\sigma_{1,1}) \sin ( \Omega_p t) \nonumber \\
&-& {\rm Im}
(\sigma_{1,1})  \cos ( \Omega_p t) ] dr \ ,
\end{eqnarray}
\begin{eqnarray}
y_{d, \rm out} &=& \frac{\pi}{M_{d, \rm out}}  \int_{r_{\rm
out,1}^{\prime}}^{r_{\rm out,2}^{\prime}} r^{\prime 2}  [ {\rm Re}
(\sigma_{1,1}) \sin ( \Omega_p t) \nonumber \\
&-& {\rm Im} 
(\sigma_{1,1})  \cos ( \Omega_p t) ] dr^{\prime}
+ r_a \sin (\Omega_p t) \ . \ \ \ \ \
\end{eqnarray}
The coordinates origin is set to the star, and the planet's position is
$(x_p, y_p) = (r_p \cos (\Omega_p t), \ r_p \sin (\Omega_p t))$.
The center of mass of the entire system, which does not move in an
inertial coordinate system, is
\begin{equation}
x_{\rm CM} =
\frac{M_p x_p + M_{d,\rm in} x_{d, \rm in} + M_{d,\rm out} x_{d, \rm
out}}{M_s + M_p + M_{d, \rm in}+ M_{d, \rm out}} \ ,
\end{equation}
\begin{equation}
y_{\rm CM} =
\frac{M_p y_p + M_{d,\rm in} y_{d, \rm in} + M_{d,\rm out} y_{d,\rm out}}
{M_s + M_p + M_{d, \rm in}+ M_{d, \rm out}} \ .
\end{equation}

The disk contribution to the stellar wobble is negligibly small
compared to the planet's direct effect.
Only the $m=1$ perturbation shifts the center of mass of the disk, and
the amplitude of the $m=1$ wave is not large enough to make a
significant movement.
There is no $m=1$ LR in the inner disk, and no significant $m=1$ wave is
excited there.
For the outer disk the $m=1$ wave is excited at the outer LR
($r_{L,1}=1.59 r_p$), which 
however is not very close to the planet, and only a relatively weak wave
is excited.
Although the wave amplitude is proportional to the planet's mass, the
amplitude of stellar wobble by the planet itself is also proportional
to the mass. Thus, the relative importance of the disk to the planet
does not change for different planet's masses.

Figure \ref{fig:cmass} shows the motion of the star.
The motion induced both by the planet and by the
disk (solid line) does not differ significantly from the motion induced
only by the planet (dashed line).
In this figure, we assumed that the disk mass is 10 times the typical
model. The disk mass inside 25AU is $0.16 M_\sun$, which is probably
close to the upper limit of real disks.
Even such a massive disk has a minor contribution to the stellar wobble.
Because the disk contribution is proportional to the disk mass,
it is negligible for most of actual disks.


\section{Wobble of the Photo-Center}

Although a disk does not perturb the stellar position significantly, its
presence may 
obstruct the precise measurement of the stellar position.
In a realistic observation, a position of a star is measured using a
relatively large beam that covers the entire star-disk system.
The mean location of the light from both the star and the disk is
interpreted as the position of the former.
If the disk is distorted and its light has any asymmetry, it may
shift the apparent position of the star.
In this section, we calculate the motion of the photo-center.
The photo-center is the mean location of a star-disk system that is
weighted by its surface brightness.
In our problem, the central star and the disk contribute to the photo-center,
while the planet's radiation is neglected.
A Jupiter mass planet is $10^{-5} \ L_{\sun}$ at the age of $10^6$ yr
and $3 \times 10^{-6} \ L_{\sun}$ at $10^7$ yr, and a 10 $M_J$ planet is 
$\sim 10^{-3} \ L_{\sun}$ at the age of $10^6-10^7$ yr (Burrows et
al. 1997).
The ratio of the luminosities of the planet and the star, $L_p/L$, is
much smaller than their mass ratio, $M_p/M$.
When a circum-planetary disk forms, its emission may contribute
(Lubow, Seibert, \& Artymowicz 1999; Tanigawa, \& Watanabe
2002; Bate et al. 2003; D'Angelo, Henning, \& Kley 2003).
We neglect the circum-planetary disk's contribution, assuming that most
of the disk material has accreted to the planet.

\subsection{Scattered Light form the Disk \label{sec:sclight}}

In our calculations of the disk radiation, we take the following
assumptions for simplicity.
We consider optical light images of the disks.
At the optical wavelengths, the disk radiation is dominated by the
scattered light of the central star, and the disk thermal emission is
neglected.
In addition, the disk is treated to be completely opaque.
Incident starlight cannot penetrate into the disk surface, but it is
completely scattered (i.e., the albedo is unity).
Thus, when the disk surface waves, only the regions that face directly to
the star shine, making the disk striped.
Figure \ref{fig:height} shows a section of a disk in the $rz$-plane.
The surface parts that are illuminated by the star are indicated by circles.
The other parts are assumed to be totally dark so they do not scatter
any light.
These assumptions make the model disks have more contrasts 
than actual disks.
In real disks, starlight can penetrate somewhat
inside the surface, and the stripes do not have clear boundaries between the
light and dark parts,
making the contrast lower.
In addition, the assumption that the incident starlight is totally
scattered at the disk surface without any absorption also works to
overestimate the disk's contrast.
Hence, the motions of the photo-centers of model disks is more sensitive to the
perturbations than actual.
The calculations below should be considered to give an upper limit of
the photo-center shift.

\subsubsection{Variation in the Disk Thickness \label{sec:thickness}}

Fluctuation of the disk surface is calculated from the results of
the wave pattern in the previous sections.
We assume that the disk always keeps hydrostatic equilibrium in the
vertical direction.
For the inner disk,
\begin{eqnarray}
&-& \frac{GMz}{(r^2+z^2)^{3/2}} \nonumber \\
&-& \frac{G M_p z}{[r^2+r_p^2 -2 r r_p \cos
 (\theta-\Omega_p t) + z^2]^{3/2}} \nonumber \\
&-& \frac{1}{\rho} \frac{dp}{dz} = 0 \ ,
\end{eqnarray}
where $\rho$ is the gas density, and $p$ is the three-dimensional gas
pressure.
We assume $z \ll r$.
In addition, outside the Roche lobe ($|r-r_p| \gg (M_p / M)^{1/3} r_p$),
the gravity of the planet (the second term) is negligible compared to
the gravity of the star (the first term).
In our model disks, the opening gap clears out the gas around the
planet's orbit, and there remains little gas inside the Roche lobe (we
ignore the circum-planetary disk).
The above equation reduces to
\begin{equation}
- \frac{G M z}{r^3} - \frac{1}{\rho} \frac{dp}{dz} = 0 \ .
\end{equation}
We use the adiabatic equation of state, $p = K_3
\rho^{\gamma_3}$.
Then, we have the density profile in the $z$-direction
\begin{equation}
\rho = \rho_{\rm mid} \left[ 1 - \left( \frac{z}{h} \right)^2
\right]^{1/(\gamma_3-1)} \ ,
\label{eq:den-z}
\end{equation}
where the subscript ``mid'' means the value at the disk midplane, and
the disk half-thickness is 
\begin{equation}
h(r, \theta)=\frac{1}{\Omega(r)}\sqrt{\frac{2 \gamma_3}{\gamma_3-1}
 \frac{p_{\rm mid}(r, \theta)}{\rho_{\rm mid}(r, \theta)}}
 = \frac{1}{\Omega(r)}\sqrt{2 \eta_{\rm mid}(r, \theta)} \ .
\label{eq:h-3d}
\end{equation}
This equation relates the disk half-thickness to the midplane density and the
pressure, although in the previous sections we calculated only the
two-dimensional values of the density and the pressure.
It is shown in the Appendix A that a similar relation,
\begin{equation}
h(r, \theta)=\frac{1}{\Omega(r)}\sqrt{\frac{2 \gamma}{\gamma-1}
 \frac{P(r, \theta)}{\sigma(r, \theta)}}
 = \frac{1}{\Omega(r)}\sqrt{2 \eta(r, \theta)} \ ,
\label{eq:h-2d}
\end{equation}
holds for the two-dimensional values.

The variables of the outer disk are defined on the inertial coordinates
$(r^{\prime}, \theta^{\prime}, z)$.
The hydrostatic equilibrium is expressed as
\begin{eqnarray}
&-&\frac{GMz}{(r^{\prime 2}+ r_a^2 - 2 r^{\prime} r_a \cos ( \theta -
 \Omega_p t - \pi) + z^2)^{3/2}} \nonumber \\
&-& \frac{G M_p z}{[r^{\prime 2}+r_b^2 -2
 r r_b \cos (\theta-\Omega_p t) + z^2]^{3/2}} \nonumber \\
&-& \frac{1}{\rho}
 \frac{dp}{dz} = 0 \ .
\end{eqnarray}
For the outer disk, $r^{\prime} \gg r_a$.
We assume $r^{\prime} \gg z$, then outside the Roche lobe
($|r^{\prime}-r_b| \gg (M_p / M)^{1/3} r_b$),
the above equation reduces to
\begin{equation}
-\frac{G M z}{r^{\prime 3}} - \frac{1}{\rho} \frac{dp}{dz} = 0 \ .
\end{equation}
Then, we have an expression of $h$ similar to that for the inner disk,
\begin{equation}
h(r^{\prime}, \theta^{\prime})=\frac{1}{\Omega(r^{\prime})}\sqrt{\frac{2
 \gamma}{\gamma-1} 
 \frac{P(r^{\prime}, \theta^{\prime})}{\sigma(r^{\prime}, \theta^{\prime})}}
 = \frac{1}{\Omega(r^{\prime})}\sqrt{2 \eta(r^{\prime}, \theta^{\prime})} \ ,
\end{equation}
where $\Omega^2(r^{\prime}) = GM/r^{\prime 3}$.

A disk height variation induced by a 10 Jupiter mass planet is shown in
Figure \ref{fig:height}.
We assume that the disk has an opaque surface at $h$, and starlight
is totally scattered at the surface.
Only the parts that are illuminated directly by the star emit radiation,
while the other parts are totally dark.

\subsubsection{Scattered light flux from the disk}

We neglect the star's diameter. 
The starlight emerges from the origin. 
If there is no obstacle on the ray to a part of the disk surface, that
part is illuminated by the star.
The parts marked by circles in Figure \ref{fig:height} are illuminated.
The starlight is assumed to be totally scattered at the disk surface
$(r, \theta, h)$.
Though the actual scattering occurs at the photo-surface of constant
optical depth, which is below the disk surface,
we approximate the photo-surface by the disk surface.
In the Appendix B, this approximation is examined.
Consider an illuminated part at $(r,\theta, h)$ on the inner disk.
An area $r dr d \theta$ on the $xy$-plane (mid-plane) corresponds to an
area of the disk surface $dS = r dr d \theta /cos \varphi$, where
$\varphi$ is an angle between  the normal vector {\boldmath $n$} of the
surface and the $z$-axis.
The surface $dS$ receives a stellar energy flux 
\begin{equation}
d L_d = L \frac{\cos \alpha dS}{4 \pi r^2} \ ,
\end{equation}
where $L$ is the stellar luminosity and $\alpha$ is the angle between
the normal vector {\boldmath $n$} and the line from $dS$ to the star.
To derive $\cos \alpha$, we assume that the waves wind so tightly that
$dh / d\theta$ is neglected comparing to $dh/dr$, i.e.,
the toroidal component of the normal vector {\boldmath $n$} is neglected
and {\boldmath $n$} $= (n_r, n_{\theta}, n_z) \approx (-\sin \varphi, 0,
\cos \varphi)$.
We then have
\begin{equation}
\cos \alpha = \sin \beta \sin \varphi - \cos \beta \cos \varphi \ ,
\end{equation}
where $\beta$ is the angle between the ray of starlight to $dS$
and the $z$-axis.
For simplicity, we assume the scattering at the disk surface is
isotropic.
The intensity of the scattered light is
\begin{equation}
I = \frac{d L_d}{\pi dS} = \frac{L \cos \alpha}{4 \pi^2 r^2} \ .
\end{equation}
An observer looks at the disk with a viewing angle $v$ ($v=0$ at
face-on), using a detector that has a solid angle $d \Omega$ from the
disk.
We define the angle $\delta$ between the line of sight (pointing to the
observer) and the normal vector {\boldmath $n$}, which satisfies
\begin{equation}
\cos \delta = - \sin v \sin \varphi \cos \theta + \cos v \cos \varphi \ .
\end{equation}
The scattered light flux from the surface element $dS$ to the detector
is then 
\begin{eqnarray}
dF_d & = & I \cos \delta dS d \Omega \nonumber \\ 
 & \approx & \frac{L}{4 \pi^2 r^2} 
\left[ 1 + \left( \frac{dh}{dr} \right)^2 \right]^{-1/2}
 \left( \frac{dh}{dr} - \frac{h}{r} \right)  \nonumber \\
& \times &
 \left( \cos v - \frac{dh}{dr} \sin v \cos \theta  \right)
 r dr d \theta d \Omega \ ,
\label{eq:scatflux_in}
\end{eqnarray}
where we used
\begin{equation}
\sin \beta  = \frac{r}{(r^2+h^2)^{1/2}} \approx 1 \ , \ 
\cos \beta = \frac{h}{(r^2+h^2)^{1/2}} \approx \frac{h}{r} \ ,
\end{equation}
\begin{equation}
\sin \varphi = \frac{dh}{dr} \left[ 1 + \left( \frac{dh}{dr} \right)^2
\right]^{-1/2} \ , \ 
\cos \varphi = \left[ 1 + \left( \frac{dh}{dr} \right)^2 \right]^{-1/2}
\ ,
\end{equation}
and $h \ll r$.

Next, consider an illuminated part at $(r^{\prime},\theta^{\prime}, h)$
on the outer disk.
The disk surface $dS^{\prime} = r^{\prime} dr^{\prime} d \theta^{\prime}
/cos \varphi^{\prime}$ receives a stellar energy flux
\begin{equation}
d L_d = L \frac{\cos \alpha^{\prime} dS^{\prime}}{4 \pi r^2} \ ,
\end{equation}
where $r$ is the distance between the disk surface element $dS^{\prime}$
and the star, and
\begin{equation}
\cos \alpha^{\prime} = \sin \beta \sin \varphi^{\prime} - \cos
\beta \cos \varphi^{\prime} + O \left[ \left( \frac{r_a}{r}
\right)^2 \right] \ ,
\end{equation}
\begin{equation}
\sin \beta = \frac{r}{(r^2+h^2)^{1/2}} \approx 1 \ , \ 
\cos \beta = \frac{h}{(r^2+h^2)^{1/2}} \approx \frac{h}{r} \ ,
\end{equation}
\begin{equation}
\sin \varphi^{\prime} = \frac{dh}{dr^{\prime}} \left[ 1 + \left(
\frac{dh}{dr^{\prime}} \right)^2 \right]^{-1/2}  \ , \ 
\cos \varphi^{\prime} = \left[ 1 + \left( \frac{dh}{dr^{\prime}}
\right)^2 \right]^{-1/2} \ .
\end{equation}
The intensity of the scattered light from $dS^{\prime}$ is
\begin{equation}
I = \frac{d L_d}{\pi dS^{\prime}} = \frac{L \cos \alpha^{\prime}}{4
\pi^2 r^2} \ ,
\end{equation}
The scattered light flux is
\begin{eqnarray}
dF_d & = & I \cos \delta^{\prime} dS^{\prime} d \Omega \nonumber \\ 
 & \approx & \frac{L}{4 \pi^2 r^2} 
\left[ 1 + \left( \frac{dh}{dr^{\prime}} \right)^2 \right]^{-1/2}
 \left( \frac{dh}{dr^{\prime}} - \frac{h}{r} \right) \nonumber \\
& \times &
 \left( \cos v - \frac{dh}{dr^{\prime}} \sin v \cos \theta^{\prime}  \right)
 r^{\prime} dr^{\prime} d \theta^{\prime} d \Omega \ ,  \ \ \ \ \
\label{eq:scatflux_out}
\end{eqnarray}
where
\begin{equation}
\cos \delta^{\prime} = - \sin v \sin \varphi^{\prime} \cos
\theta^{\prime} + \cos v \cos \varphi^{\prime} \ .
\end{equation}
A simulated image of a face-on disk is shown in Figure \ref{fig:image_std_i0}.

\subsection{Motion of the Photo-Center}

\subsubsection{Face-on Disks}

If a disk is observed at face-on ($v=0$), then the most important
mode in moving the photo-center is the $m=1$ wave, which also shifts the
mass center of the disk. 
Modes of $m \ge 2$ can also shift the photo-center slightly, because the
disk emission is a non-linear function of the perturbations (see
eqs. [\ref{eq:scatflux_in}] and [\ref{eq:scatflux_out}]) and does not
have a symmetry after summing up $m \ge 2$ perturbations.
The contributions of $m \ge 2$ modes are, however, small.

Figure \ref{fig:cl_std}$a$ shows the motion of the photo-center of the
star-disk system.
The disk scattered light significantly affects the motion of the photo-center,
which is shown by the solid line.
The motion of the disk partly cancels the stellar motion, making the
amplitude of the photo-center's motion about two-third of that without the disk
(shown by the dashed line). 
The disk contribution to the motion of the photo-center is significant.
In this figure the planet's mass is a Jupiter mass ($M_J \approx 10^{-3}
M_{\sun}$). 
For a larger mass planet, however, the disk's relative contribution
becomes smaller.
Figure \ref{fig:cl_std}$b$ shows that for a 10 $M_J$ planet the motion
of the photo-center with a disk does not significantly differ from the motion
without the disk.
The amplitude of the star's dynamical motion is proportional to the
planet's mass, while the disk contribution to the motion of the photo-center is
not proportional to it.
The contrast of the spiral pattern shown in Figure
\ref{fig:image_std_i0} is nearly saturated even though the planet is as
small as Jupiter.
Thus, a further increase in the planet's mass does not cause a
significant increase in the spiral pattern contrast nor a much larger shift of
the photo-center by the disk.
A more massive planet results in a decreasing importance of
the disk in the photo-center's motion.

\subsubsection{Inclined Disks}

Waves of $m=2$ and higher do not move the center of mass of a
disk, and only slightly shift the photo-center of a face-on disk.
If a disk is observed at an inclined angle, however, $m \ge 2$ waves
also contribute in moving the photo-center. 
When a flared disk is inclined to the observer, the half part 
on the farther side is illuminated more efficiently than the other half.
This imbalance breaks the symmetry of $m \ge 2$ waves.

Figure \ref{fig:image_std_inc} shows the scattered light image of a disk
that is $75^{\circ}$ inclined to the line of sight.
The bright part of the disk is the farther side to the observer, and the
closer side is dark.
The position of the photo-center shifts from the star to the bright side.
This shift is as large as the disk radius multiplied by the disk-to-star
luminosity ratio, $r_d L_d/L$, which is much larger than the amplitude
of the stellar wobble.
In observations of the stellar position, however, it leads to only an
additional offset to the mean stellar 
position, because observational beam sizes are usually larger than the
disk radius.
An important observational quantity is the motion of the photo-center
with time.
As the planet orbits, the wave pattern rotates.
This causes a fluctuation of the photo-center, because only
a half disk is on the bright side.
Figure \ref{fig:cl_std_i60}$a$ shows the motion of the photo-center caused by a
1 $M_J$ planet.
The disk is $60^{\circ}$ inclined to the line of sight.
We show the relative motions from the average position.
The asymmetry of disk light between the bright part and the dark part
causes a large offset of the photo-center position from the stellar
position, but this offset is subtracted in plotting Figure
\ref{fig:cl_std_i60}.
Presence of the disk changes the orbital shape of the photo-center in addition
to enlarging the motion.
The amplitude of the disk effect is similar to that in the face-on case.
Hence, the disk light interferes with precise determination of the
planet's orbit.
For more massive planets, the relative importance of a disk becomes
smaller.
Figure \ref{fig:cl_std_i60}$b$ shows that if the planet's mass is as
large as 10 $M_J$, the disk effect is not very significant.

\subsection{Various Models}

\subsubsection{Low Viscosity}

In our typical models, we used $\alpha_{\rm vis} = 10^{-2}$ for the viscosity
parameter.
Here, we consider a model with a lower viscosity $\alpha_{\rm vis} = 10^{-3}$.
As shown in Figure \ref{fig:damp}, waves have attenuated before arriving
at $r_{\rm out,2} = 5 r_p$ for an $\alpha_{\rm vis} = 10^{-2}$ disk.
In a low viscous disk, however, waves can reach as far as $10 r_p$, so we
use a disk model with $r_{\rm out,2} = 10 r_p$.
Figure \ref{fig:cl_a1D-3} shows the motion of the photo-center of a $60^{\circ}$
inclined disk.
The motion of the photo-center has a different orbital shape from that of the high
viscous disk, though their amplitudes are similar.
The wave spiral pattern in a lower viscous disk extends to a larger region
of the disk, leading to a different shift of the photo-center of the disk.
This result means that the motion of the disk's photo-center is sensitive to
the wave pattern, which changes according to the disk properties.
Thus, if the disk contamination to the photo-center is significant, it is
difficult to remove this contamination.
 
\subsubsection{Narrow Gap}

In our typical models, we assume a gap between $[0.6r_p, 1.4r_p]$ (where
the density is below half of the value without the planet).
If the planet's mass is larger than the Jupiter mass, it clears the disk gas
from the $m=2$ inner LR at $0.63r_p$ to the $m=2$ outer
LR at $1.31 r_p$ (Lin \& Papaloizou 1986; Takeuchi et al. 1996).
Thus, the gap size assumed here is the typical value, but we also made
an experiment to see what occurs if the gap is narrower.
The motion of the photo-center in Figure \ref{fig:cl_ngap} is calculated
with an assumed gap between $[0.8r_p, 1.2r_p]$.
The narrow gap allows wave excitation of modes as high as $m \approx 6$,
while in the wide gap case only $m \la 3$ waves are excited.
It causes a stronger disturbance in the disk.
The fluctuation of the photo-center induced by the disk is so large that
the precise measurement of the planet's orbit becomes difficult.

\section{Astrometry Shift Observed by the SIM}

The astrometric wobble in the  position of a 1 $M_{\sun}$ star  at a
distance of 140 pc due to a Jupiter mass planet in a 5 AU orbit is
$\sim 36 \ \mu$as ($\sim 7 \ \mu$as for 1 AU orbit). 
The interferometric observations by SIM will have the accuracies 
to measure such small angular displacements. However, SIM delay measurements 
correspond to the position of the photo-center. In order to derive the 
astrometry parameters of the target star, such as  parallax, proper 
motion, and star's wobble about the center of mass, we have to assume
that the measured photo-center position shifts represent the true
astrometry signal of the target. 
Often it is assumed that the target has only non-luminous  stellar 
or sub stellar companions. If  the target star is not in a clean environment, 
but embedded in a disk, it is likely that   the position of the
photo-center as measured by SIM delay is not a true representation of
the target star position.

SIM will have the ability to measure the phase of the complex visibility
of the
star with unprecedented accuracy. In the visible wavelengths the star is a 
million time  brighter than the planets.  Therefore, in the  idealized
case of a
single star with planets, the brightness distribution can be assumed to be 
point source with the photo-center coincident with the star position. The 
complex visibility of the brightness distribution has an amplitude unity with 
the phase changing as the photo-center wobbles around the star-planet
center of mass. 
The visibility phase shift $\Delta\phi$ measured by the interferometer 
with a baseline $b=10$ m is related to the angular separation $\Delta \theta$
between the center of mass and the star position:
\begin{equation}
\Delta\phi = \frac{2 \pi b}{\lambda} \Delta\theta \ ,
\label{eq:phaseshift}
\end{equation}
where $\lambda$ is the  wavelength. The measured visibility phase 
$\Delta\phi$ can result from (i) the systemic motion of the target star, (ii) 
the fringe measurement errors due to the photon noise, and (iii) from 
the brightness distribution around  the target star, such as the
scattered light
from the disk. The photon   and instrumental noises  
introduce a randomly varying position shift which can be reduced by longer 
observing time.  
However, the photo-center shift induced due to 
the brightness structures around the star introduces a 
systematic position error, which    vary with time only as the orientation of 
the baseline changes. This makes interpretation of the measured astrometry 
signal in terms of wobble induced by planet(s) difficult. It can be partly 
removed  by modeling if we know the target environment well enough.
 
The disk's contamination to the measured astrometric shift is especially
significant if the planet's mass is smaller than about 10 Jupiter mass. 
An interferometer is not too sensitive to structures on a much larger
spatial scale than the fringe size $\lambda/b$. 
In the case of an extended source with a uniform brightness the
interferometer fringe visibility decreases with the increasing source
size $\alpha_{\rm src}$ as sinc($\pi \alpha _{\rm src} \lambda /b $),
where ${\rm sinc} \ x$ is $(\sin x) / x$.
It becomes zero when the source size is equal to the fringe size. 
For SIM the fringe  size is  about 10 mas  at 0.5$\micron$.
Although an extended disk with a radially smooth symmetric distribution
(with size $\gg$ 10 mas) will be  well within the field view of SIM, its
fringes will be washed out significantly and therefore, will not
contribute to the fringe visibility of the target star.
On the other hand, any small scale structures (size $<$ 10 mas) present
in the scattered light within the SIM field of view, such as those
caused by disk spirals or gaps, can produce fringes. 
Thus, the small scale structures in the disk can
contribute to the complex fringe visibility of the target star, resulting
in a phase shift that will be interpreted as astrometry signal.

In this section, we present the results of the simulated  astrometry
shifts for some of the disk light models. 
In order to simplify the computations we assumed 8 spectral channels covering 
the wavelength range 400 to 900 nm. The complex visibility phases of the 
star-disk structure were computed for each channel for given orientation
of the baseline.
We use SIM parameters $b = 10$ m and mirror size 33 cm.  
Using equation (\ref{eq:phaseshift})
we estimate the astrometric  position shift for each channel. The final values 
for astrometry shifts were obtained by averaging over the channels weighted by 
SIM sensitivity in each channel and by the stellar photon flux  in each
channel.  
The stellar magnitudes, spectral type, and estimated SIM throughput were
used to compute the number of photons detected in each spectral channel.
For SIM throughput we adopted the values obtained from the design team
for the aperture, transmission, and detector efficiencies. 
We also assumed a field stop of $1 \arcsec$, but
the band width smearing resulting from  combining the astrometry shifts
over all channels reduces significantly the effective field of view to
30 mas.
 
For these simulations we assumed the objects are at Taurus distance, 140 pc. 
A Jupiter mass planet orbits at 5 AU from a solar-mass star, leading to a
motion of the star with an amplitude $36 \ \mu$as.
The stellar photospheric emission is approximated by a blackbody of 4000
K.
The disk scattered light image is constructed on the numerical grids of $2001
\times 2001$.
The star is set on the central grid at $(1001,1001)$.
Figure \ref{fig:sim_i0} shows simulated astrometry shifts for the SIM.
We show only the additional shift caused by a disk, by subtracting
dynamical shifts induced by a planet.
A disk of our typical model is viewed at face-on.
The amplitude of the measured variation in the photo-center's position is less
than 1 $\mu$as, which is much smaller than the amplitude of the stellar
motion, $36 \ \mu$as.
Figure \ref{fig:sim_i60} shows the astrometry shifts for an inclined disk.
The disk is observed at an inclination $v=60^{\circ}$.
We show the shifts measured with two baselines of the SIM. One is
parallel to the major axis of the disk image, and the other is perpendicular.
The amplitudes of the measured variations in the photo-center's position
are less than 1 $\mu$as.
Hence, presence of a disk does not make a considerable contamination
on the measurement of the stellar position.
Figure \ref{fig:sim_rp10_i60} shows a case in which a planet orbits at 10 AU,
which is twice of the typical models.
The disk size is set twice, 50 AU. The size of wave pattern also becomes
twice, and we expect the amplitude of the photo-center's motion is
two times larger.
The predicted astrometry shifts are however comparable to the case of a
planet at 5 AU.
The scale length of the wave pattern is two times larger, and because a
larger structure is less sensitive to interferometer observations, the
disk has a smaller effect on the astrometry shift.
This decrease in the sensitivity nearly cancels the increase in the photo-center's
motion.
The dynamical shift by a planet is proportional to the
planet's orbital radius.
For a planet at 10 AU, the dynamical shift has an amplitude $71 \ \mu$as.
Hence, the measurement of a planet with a lager orbital radius suffers less
contamination of the disk light, although we need a longer observing
time to detect such a long-period planet.

\section{Conclusions}

The influence of a circumstellar disk on the motion of the star induced
by a planet is analyzed.
We show the following.

(1) The $m=1$ disturbance in a disk dynamically moves the stellar
position.
The disk dynamical effect is, however, much smaller than the planet's
gravity, and can be neglected.

(2) Disk emission moves the photo-center of a star-disk system.
In some cases, the disk considerably shifts the photo-center and
affects the precise measurement of the stellar position.
A significant contamination from the disk occurs if the planet's
mass is smaller than 10 Jupiter mass, or if the gap in the disk is narrow.
For more massive planets, the disk contamination is less important.

(3) Actual measurement of the stellar position using an
interferometer are insensitive to the extended large scale structures in
the disk surface brightnesses. 
The SIM measurement, for example, does not suffer a significant
contamination from the scattered light of the disk.

We consider that the above results provides an upper limit of the disk
light contamination, because our models assume maximum contrast
of the disk spiral pattern (see \S4). 
In addition, the dust residing above the disk scatters the
starlight and illuminates the disk surface with a large incident
angle. This secondary illumination from the upper dust also reduces the
disk's contrast.
Hence, in actual disks, the disk light contamination is probably smaller.
We need a more realistic radiative transfer calculation in order to
take account of such effects and to evaluate precise values of the disk
light contamination.

In this paper, the planet's orbit is assumed to be circular.
A planet having an eccentric orbit opens a wider gap than a planet of
a circular orbit (Artymowicz \& Lubow 1994). Because a wider gap
suppresses the wave excitation,
the disk's spiral pattern is weaker. Hence, for a planet of 
an eccentric orbit, the disk's contamination is less important.
If the planet's orbit is inclined to the disk, however, the planet excites
bending waves (Shu, Cuzzi, \& Lissauer 1983; Artymowicz 1994; Ostriker
1994), which may have a large amplitude of wavy pattern on the disk surface. 
In such cases, the disk's contamination may be more significant than that
estimated in this paper, and should be examined as a future work.
 
\acknowledgements
We thank L. Hartmann, C. Beichman, and G. Bryden for useful
conversations and P. S. Lykawka for comments.
We also thank the referee's useful comments.
This work was supported in part by an NSF grant AST 99 87417 and in part
by a special NASA astrophysical theory program that supports a joint
Center for Star Formation Studies at UC Berkeley, NASA-Ames Research
Center, and UC Santa Cruz.
This work was also supported by NASA NAG5-10612 through its Origin
program and by JPL 1228184 through its SIM program, and by
the 21st Century COE Program of Japan, ``Origin and Evolution of Planetary
Systems''.

\appendix

\section{Derivation of the Disk Thickness from the Two-dimensional
Density and Pressure}

In this appendix, we derive the disk half-thickness from the two-dimensional
values of the density and the pressure (eq. [\ref{eq:h-2d}]).
Integrating the gas density $\rho$ (eq. [\ref{eq:den-z}]) in the
$z$-direction gives the surface density,
\begin{equation}
\sigma = 2 \int_0^{\infty} \rho dz = \rho_{\rm mid} h B \left(
\frac{1}{2}, \frac{1}{\gamma_3 -1} + 1 \right) \ .
\label{eq:sden}
\end{equation}
The beta function $B$ is
\begin{equation}
B \left(\frac{1}{2}, a + 1 \right) = \frac{\Gamma(1/2)
\Gamma(a+1)}{\Gamma(a+3/2)} = 2 \int_0^1 (1 - x^2)^a dx \ ,
\end{equation}
where $\Gamma$ is the gamma function.
The two-dimensional pressure is similarly
\begin{equation}
P = 2 \int_0^{\infty} p dz = p_{\rm mid} h B \left(
\frac{1}{2}, \frac{\gamma_3}{\gamma_3 -1} + 1 \right) \ .
\label{eq:pre2}
\end{equation}
Using equations (\ref{eq:sden}) and (\ref{eq:pre2}) with the relation
\begin{equation}
B \left( \frac{1}{2}, \frac{\gamma_3}{\gamma_3 -1} + 1 \right) =
\frac{2 \gamma_3}{3 \gamma_3 -1}
B \left( \frac{1}{2}, \frac{1}{\gamma_3 -1} + 1 \right) \ ,
\end{equation}
the disk half-thickness (eq. [\ref{eq:h-3d}]) is
\begin{equation}
h=\frac{1}{\Omega}\sqrt{\frac{2 \gamma_3}{\gamma_3-1}
 \frac{p_{\rm mid}}{\rho_{\rm mid}}}
=\frac{1}{\Omega}\sqrt{\frac{3 \gamma_3-1}{\gamma_3-1}
 \frac{P}{\sigma} } \ .
\end{equation}
From equations (\ref{eq:sden}), (\ref{eq:pre2}), and $h \propto
\rho_{\rm mid}^{(\gamma_3 -1)/2}$, we see the relation between the two-
and three-dimensional $\gamma$'s as $\gamma = (3 \gamma_3-1)/(\gamma_3
+1)$.
Then, the disk half-thickness is 
\begin{equation}
h=\frac{1}{\Omega}\sqrt{\frac{2 \gamma}{\gamma-1}
 \frac{P}{\sigma} } \ ,
\end{equation}
which is identical to equation (\ref{eq:h-2d}).

\section{Location of the Photo-surface}

In \S\ref{sec:sclight}, it is assumed that the starlight scatters at the
disk surface $(r,\theta,h)$ at which the gas density drops to zero.
The realistic scattering, however, occurs at the photo-surface of
constant optical depth, which is below the disk surface.
In this appendix, we examine this assumption.
The optical depth at $(r, \theta, z)$ to the starlight is 
\begin{eqnarray}
\tau & \sim & \frac{1}{\cos \alpha} (h-z) \rho(z) \kappa \nonumber \\ 
& \sim & \frac{1}{\cos \alpha} 2^{1/(\gamma_3-1)} \kappa \rho_{\rm mid}
h \left( \frac{h-z}{h} \right)^{\gamma_3 /(\gamma_3-1)} \ ,
\end{eqnarray}
where $\alpha$ is the incident angle of the starlight, $\kappa$ is the
opacity, which is $\sim 100$ for the visible light if the dust-to-gas
ratio is similar to the solar abundance and the dust is well mixed to
the gas (Miyake \& Nakagawa 1993).
In the second equality, equation (\ref{eq:den-z}) is used. 
The condition of the photo-surface, $\tau = 1$, reduces to
\begin{equation}
\frac{h-z}{h} \sim 2^{-1/\gamma_3} \left( \frac{\cos \alpha}{\kappa
\rho_{\rm mid} h} \right)^{(\gamma_3 -1)/\gamma_3} \ .
\label{eq:ph-surface}
\end{equation}
We use the values at 10 AU: $\gamma_3 = 1.4$, $\cos \alpha \approx
(dh/dr -h/r) \sim 0.1$, $\kappa \sim 100 \ {\rm cm^2 \ g^{-1}}$,
$\rho_{\rm mid} \sim 10^{-11} {\rm g \ cm^{-3}}$, and $h \sim 10^{13}$ cm.
Equation (\ref{eq:ph-surface}) becomes $(h-z)/h \sim 0.02$.
Thus, the photo-surface almost coincides with the disk surface.
However, we are considering an era after the formation of gas giant
planets.
If most of the dust has become large bodies, the opacity $\kappa$ is
greatly reduced. In addition, the dust sedimentation to the midplane also
reduces the opacity near the disk surface.
These effects may shift the photo-surface significantly below the disk
surface, and are the subject of future investigations.



\clearpage

\begin{figure}
\epsscale{1.0}
\plotone{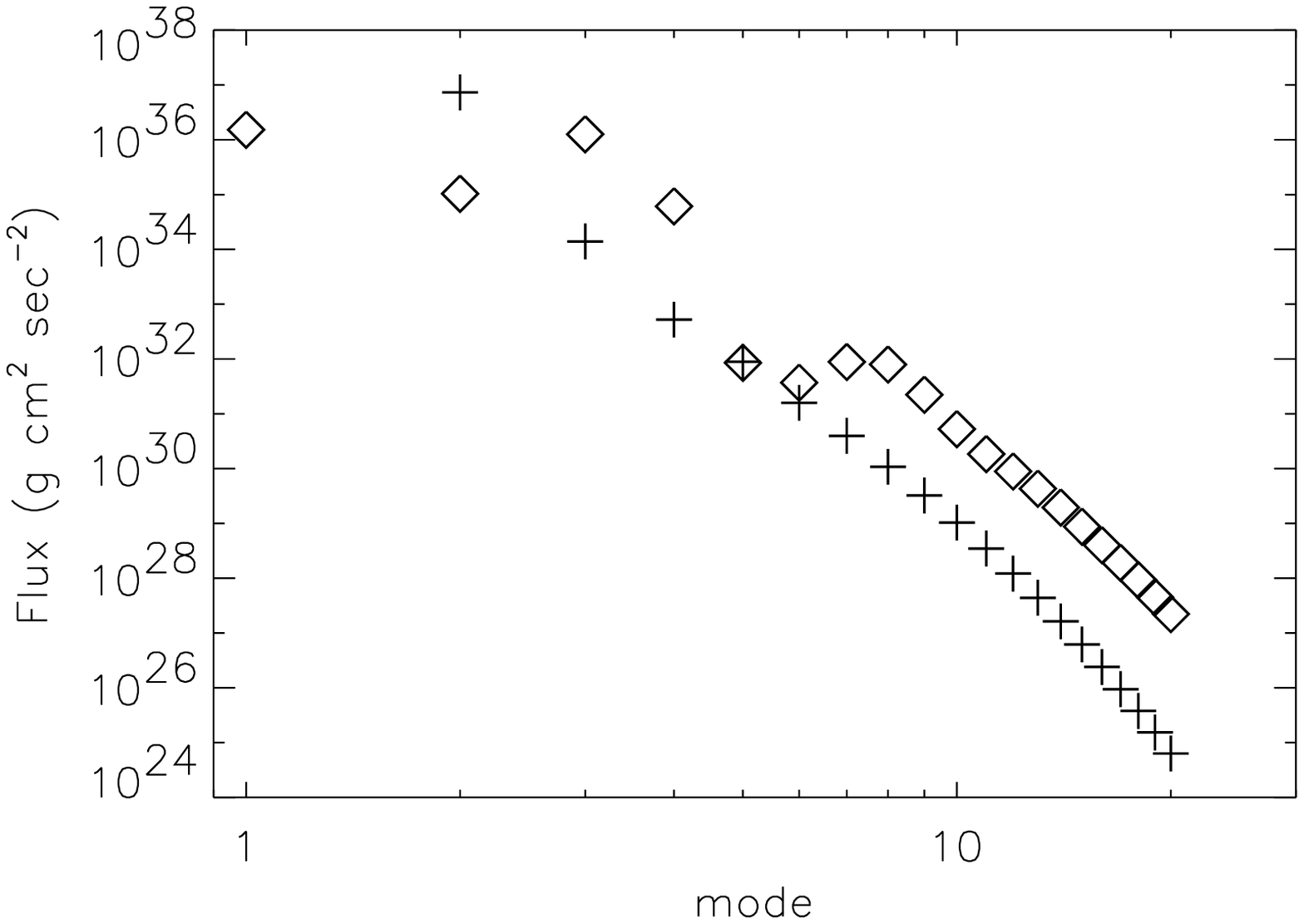}
\caption{
Angular momentum fluxes carried by the waves of various modes $m$.
The diamonds $(\Diamond)$ and crosses $(+)$ show fluxes in the
outer disk and in the inner disk, respectively.
\label{fig:flux}
}
\end{figure}

\begin{figure}
\epsscale{2.2}
\plottwo{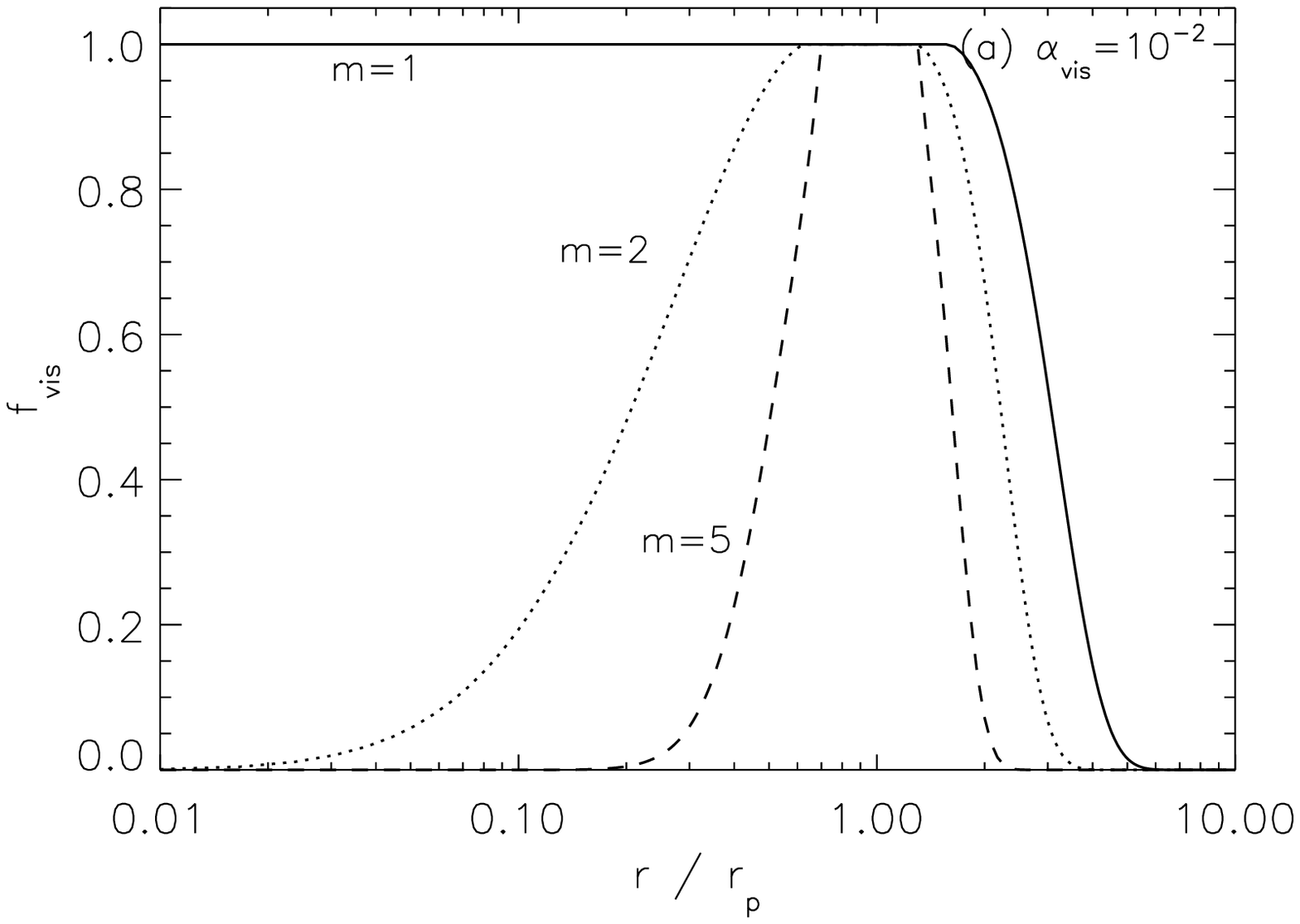}{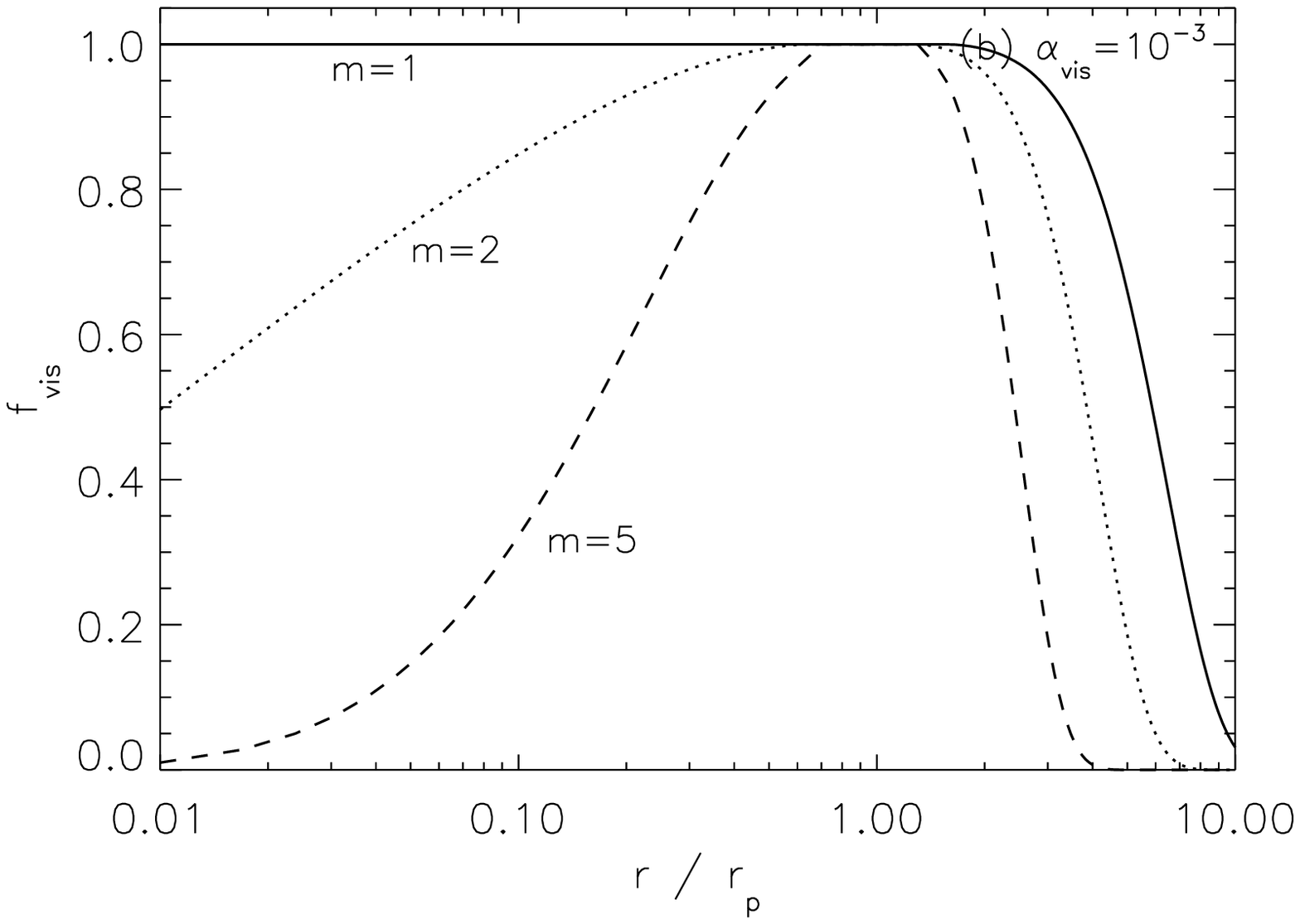}
\caption{
Wave damping factor $f_{\rm vis}$.
The solid, dotted, and dashed lines correspond to $m=1$, 2, and 5 waves,
respectively.
The damping factor (eq. [\ref{eq:fvis}]) is calculated for the inner part
inside the inner Lindblad resonances (ILRs) and for the outer part outside the
outer Lindblad resonances (OLRs).
Between the ILR and the OLR, $f_{\rm vis}$ is assumed to be unity.
For the $m=1$ wave, there is no ILR and $f_{\rm vis}$ is set to unity in
the inner disk inside the planet's orbit.
($a$) $\alpha_{\rm vis} = 10^{-2}$.
($b$) $\alpha_{\rm vis} = 10^{-3}$.
\label{fig:damp}
}
\end{figure}

\begin{figure}
\epsscale{1.0}
\plotone{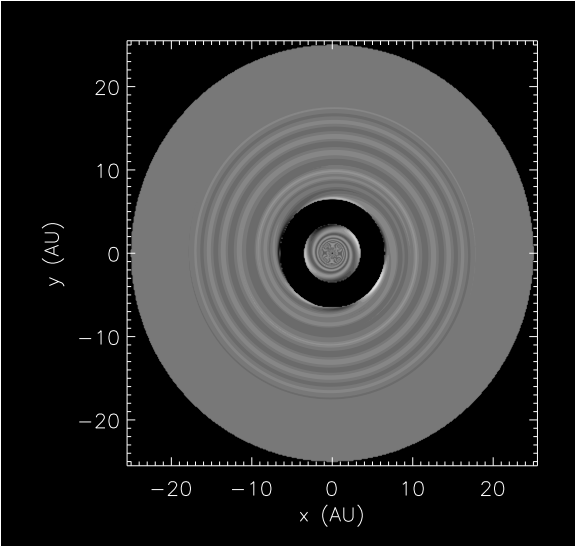}
\caption{
The surface density perturbation normalized by the unperturbed value,
$\sigma_1(r)/\sigma_0(r)$, is plotted.
A planet is at $(5,0)$, and opens a gap.
\label{fig:img_sden}
}
\end{figure}

\begin{figure}
\epsscale{1.0}
\plotone{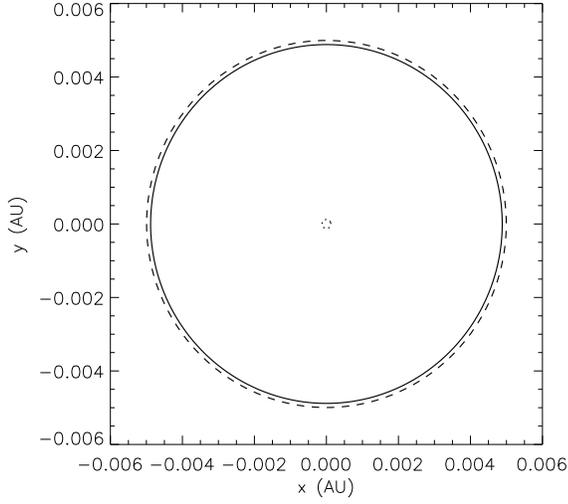}
\caption{
Movement of the stellar position in the inertial coordinates. 
A Jupiter mass planet orbits at 5 AU.
The movement caused both by the planet and the disk
is shown by the solid line.
The dashed line shows the movement caused only by the planet, ignoring
the disk's effect.
The dotted line shows the contribution of the disk (the difference
between the solid and dashed lines).
\label{fig:cmass}
}
\end{figure}

\begin{figure}
\epsscale{1.0}
\plotone{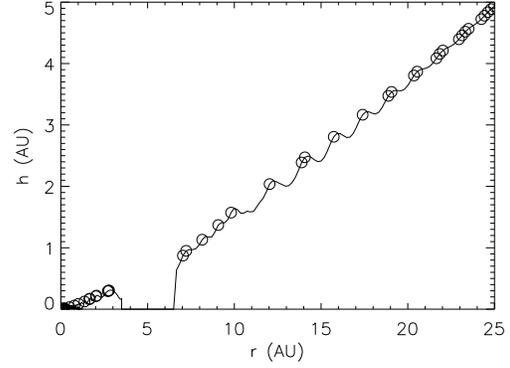}
\caption{
Fluctuation of the disk surface.
A variation in the disk half-thickness
in the planet's direction ($\theta=\Omega_p t$) is shown by the solid
line. 
The portions marked by the circles are directly illuminated by the star.
To show the fluctuation clearly, we use a model with
a 10 Jupiter mass planet ($M_p=10 M_J$) at 5 AU.
A gap opens around the planet's orbit.
\label{fig:height}
}
\end{figure}

\begin{figure}
\epsscale{1.0}
\plotone{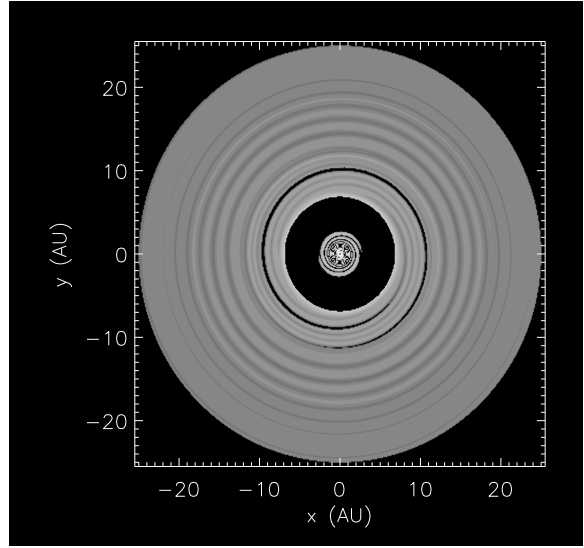}
\caption{
Simulated scattered light image of a face-on disk. 
The disk is perturbed by a Jupiter mass planet ($M_p=M_J$).
Gray scale shows logarithmic values of the disk flux.
\label{fig:image_std_i0}
}
\end{figure}

\begin{figure}
\epsscale{2.2}
\plottwo{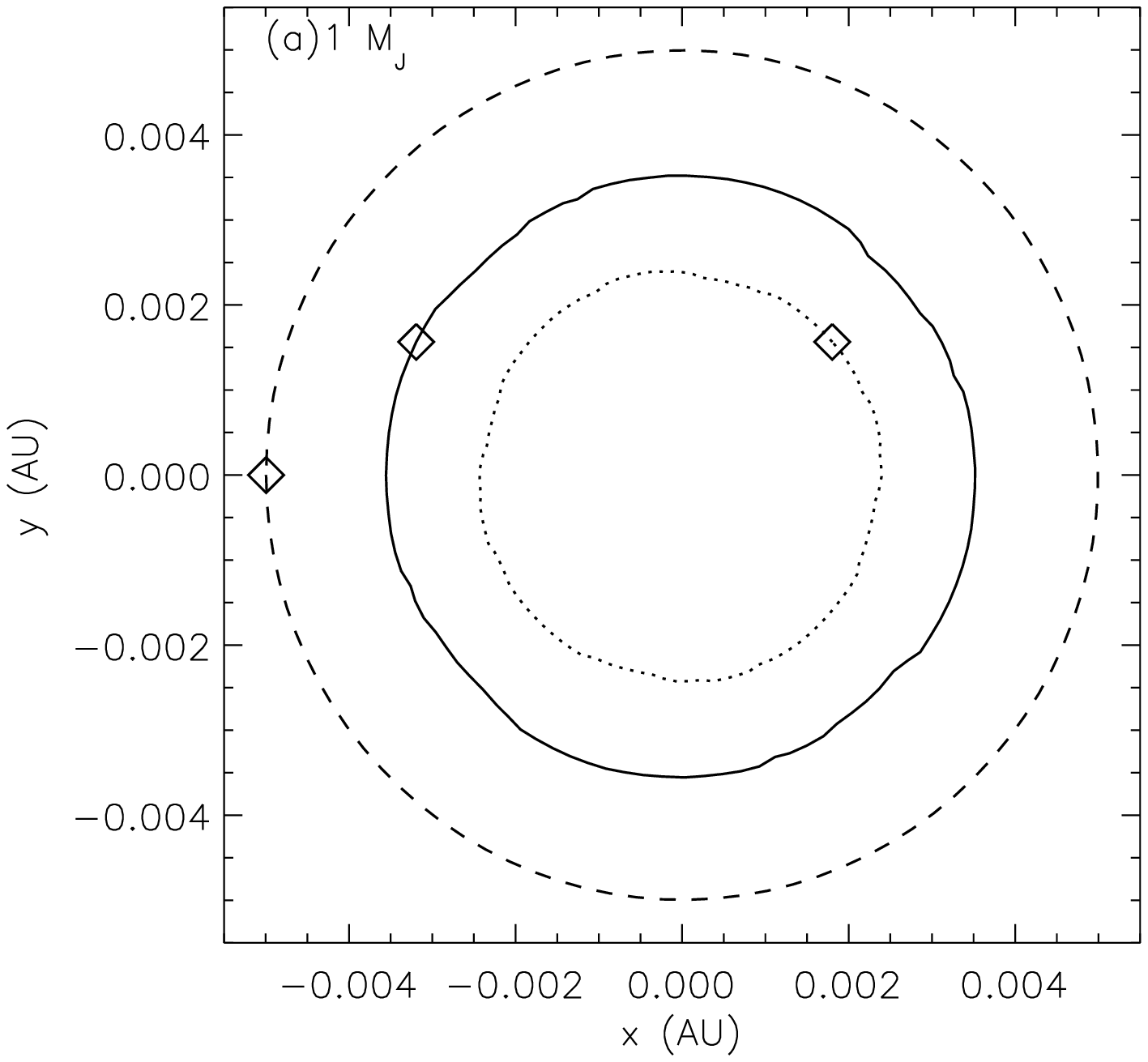}{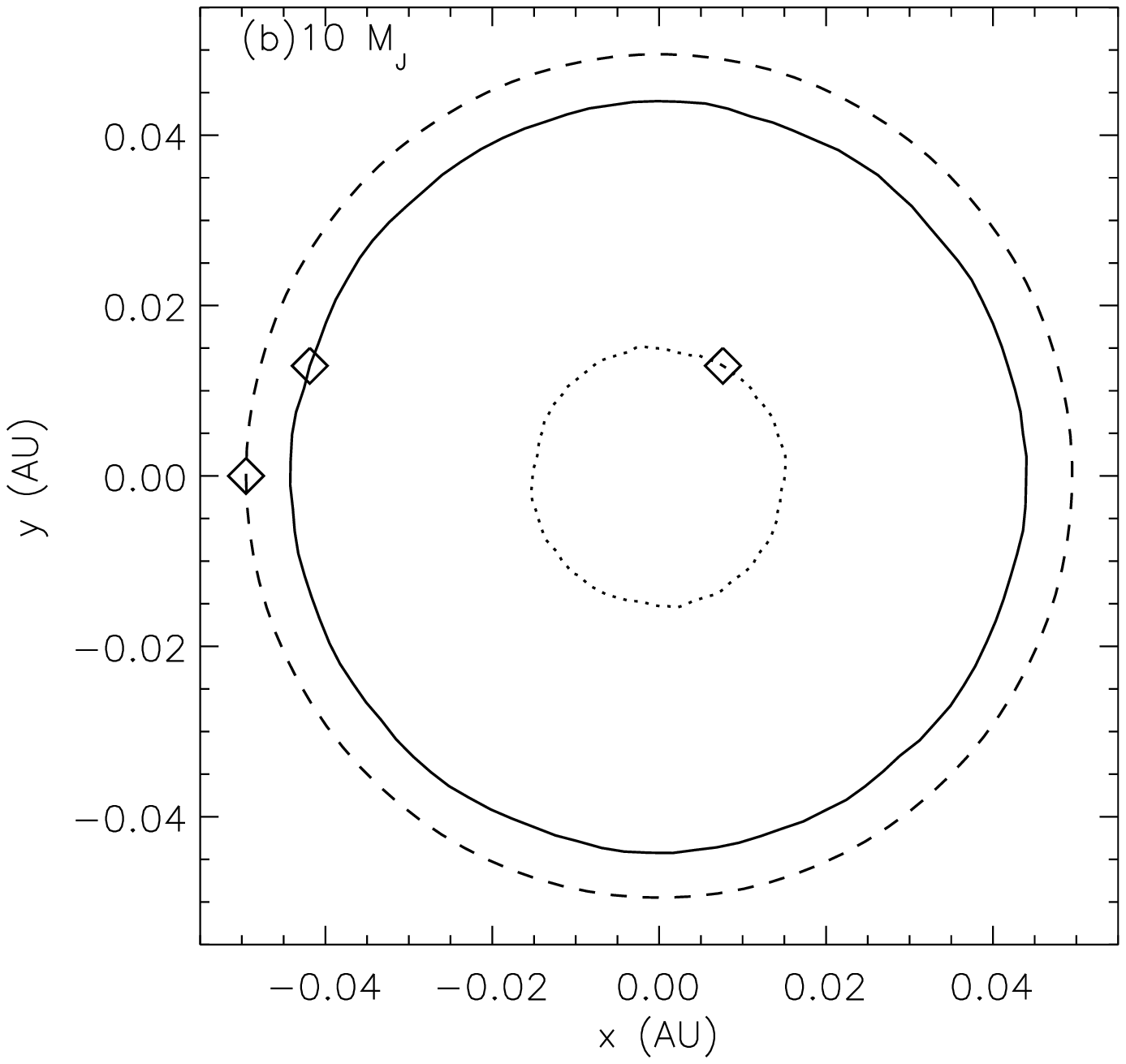}
\caption{
The shift of the photo-center for a face-on disk.
The solid line show the motion of the photo-center of the star-disk
system.
The dashed line show the motion of the star only by the planet, assuming
no disk.
The dotted line shows the disk's contribution (the difference of the
above two results).
The diamonds show the positions of the photo-centers when the orbital phase of the
planet is zero.
($a$) The photo-center motions with a Jupiter mass planet ($M_p=M_J$).
($b$) The motions with a 10 Jupiter mass planet ($M_p=10M_J$).
\label{fig:cl_std}
}
\end{figure}

\begin{figure}
\epsscale{1.0}
\plotone{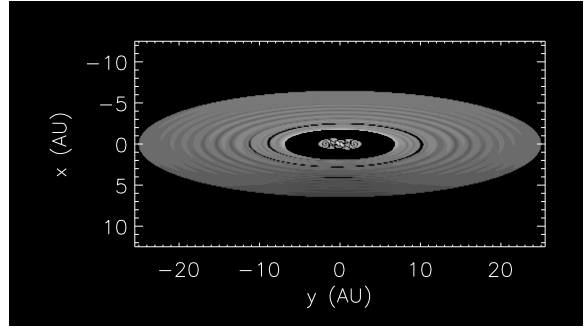}
\caption{
Simulated scattered light image of a inclined disk. 
The disk is perturbed by a Jupiter mass planet ($M_p=M_J$) and viewed
at $v=75^{\circ}$ ($v=0$ for a face-on disk).
The top half of the disk is the farther side to the observer, while the
bottom half is the closer side.
Gray scale shows logarithmic values of the disk flux.
\label{fig:image_std_inc}
}
\end{figure}

\begin{figure}
\epsscale{2.2}
\plottwo{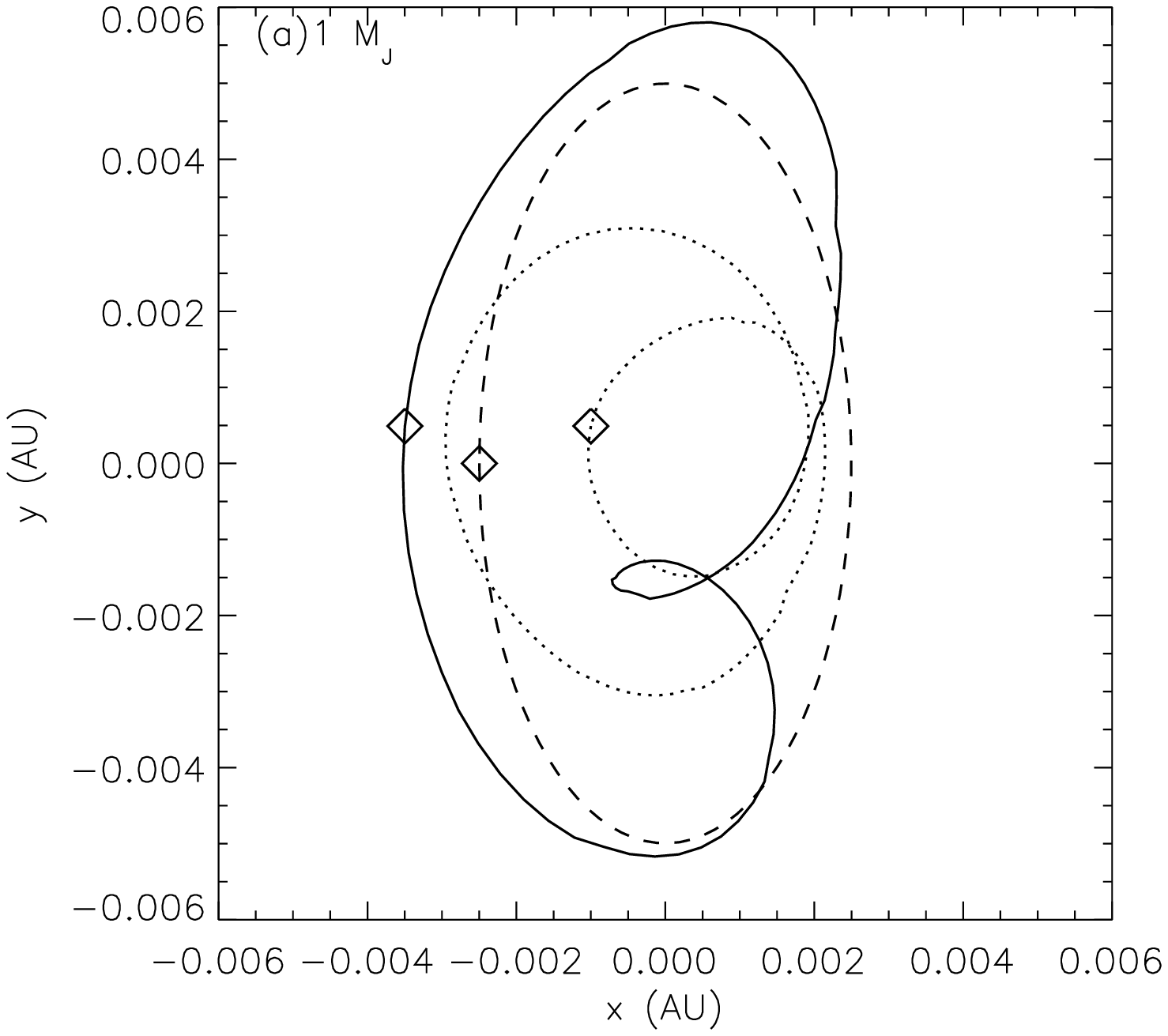}{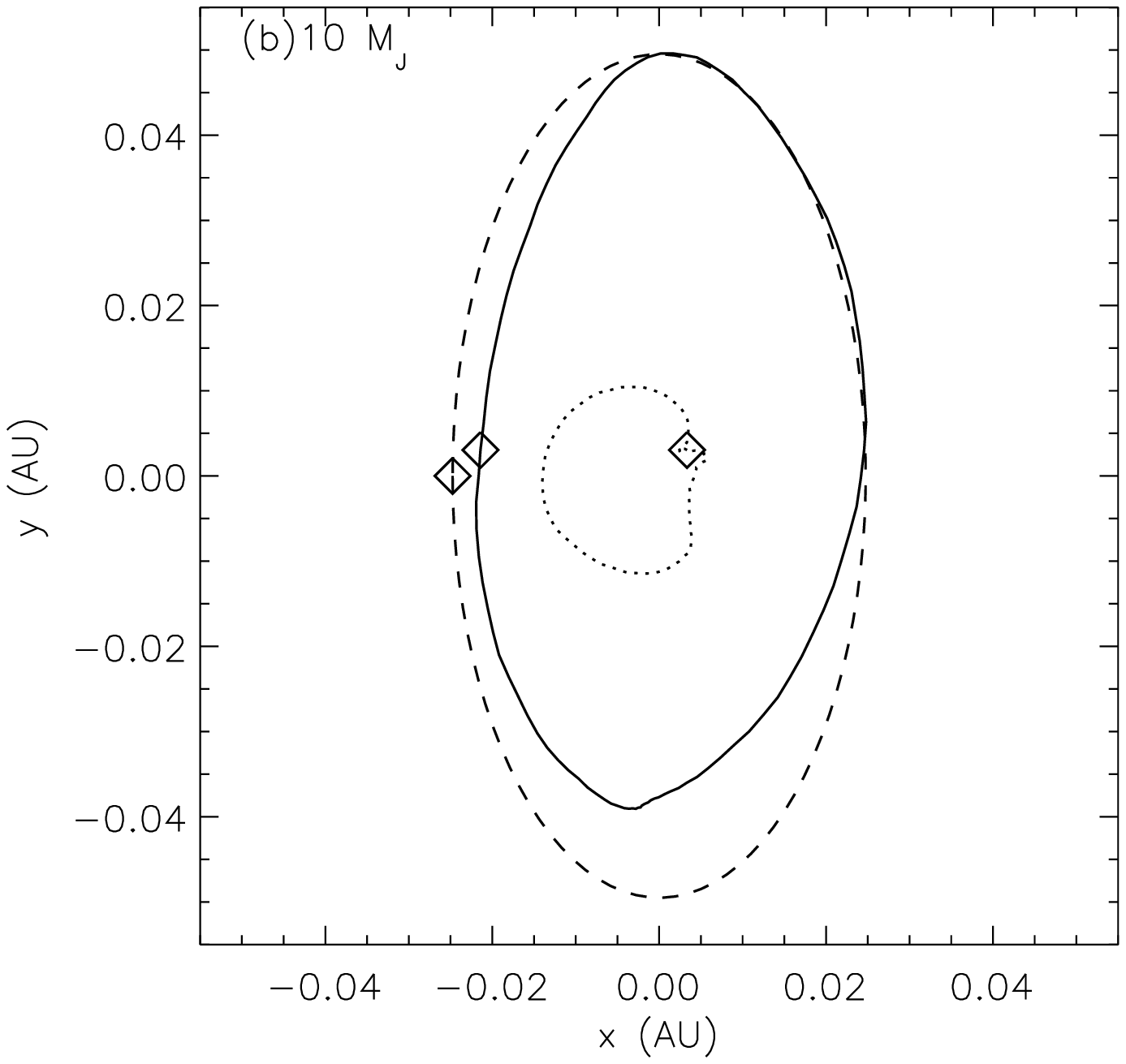}
\caption{
The shift of the photo-center.
The disk is viewed at an inclined angle $v=60^{\circ}$.
The solid, dashed and dotted lines represent shifts in photo-centers
(see Fig. \ref{fig:cl_std})
($a$) The photo-center motions with a Jupiter mass planet ($M_p=M_J$).
($b$) The motions with a 10 Jupiter mass planet ($M_p=10M_J$).
\label{fig:cl_std_i60}
}
\end{figure}

\begin{figure}
\epsscale{2.2}
\plottwo{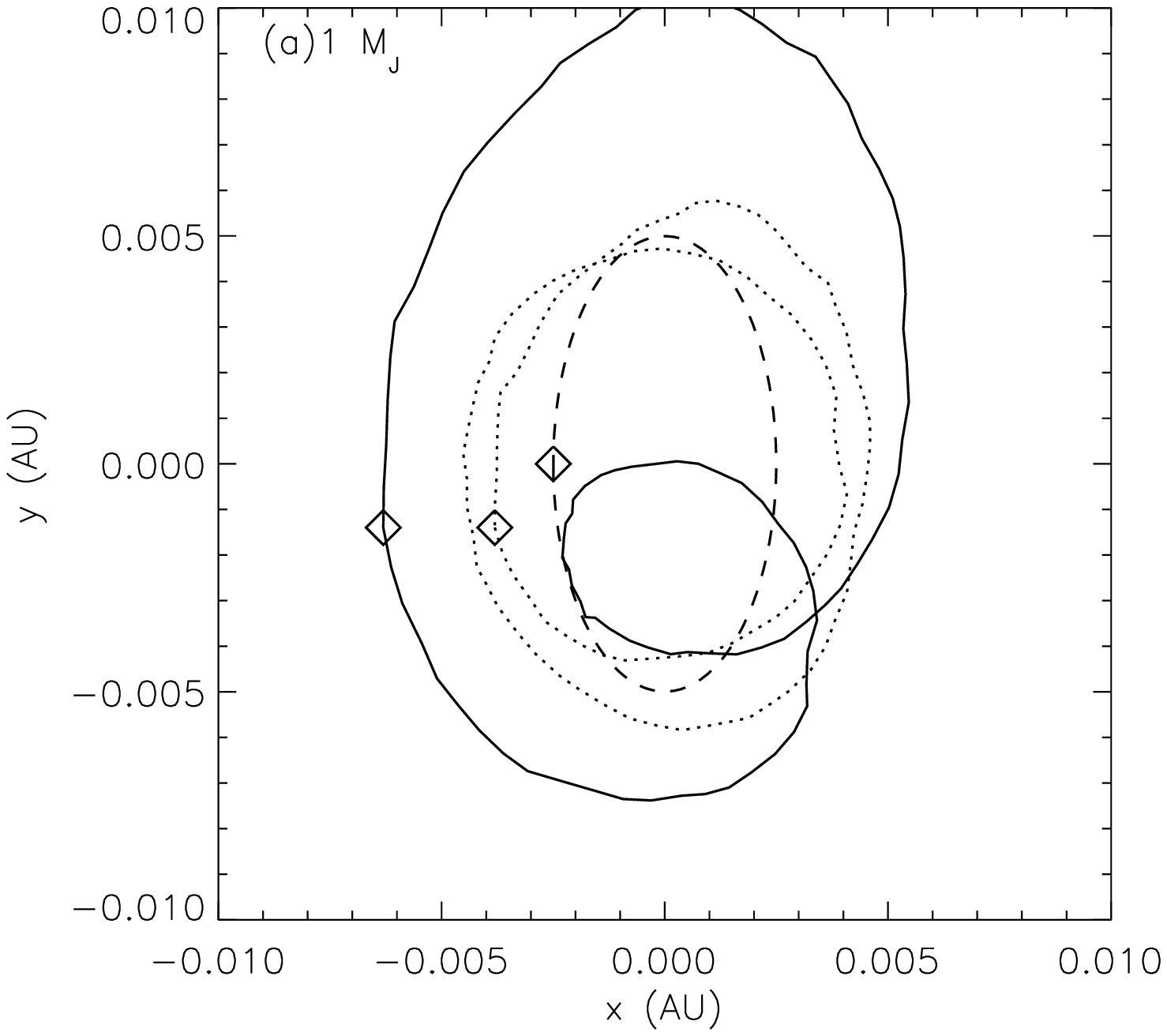}{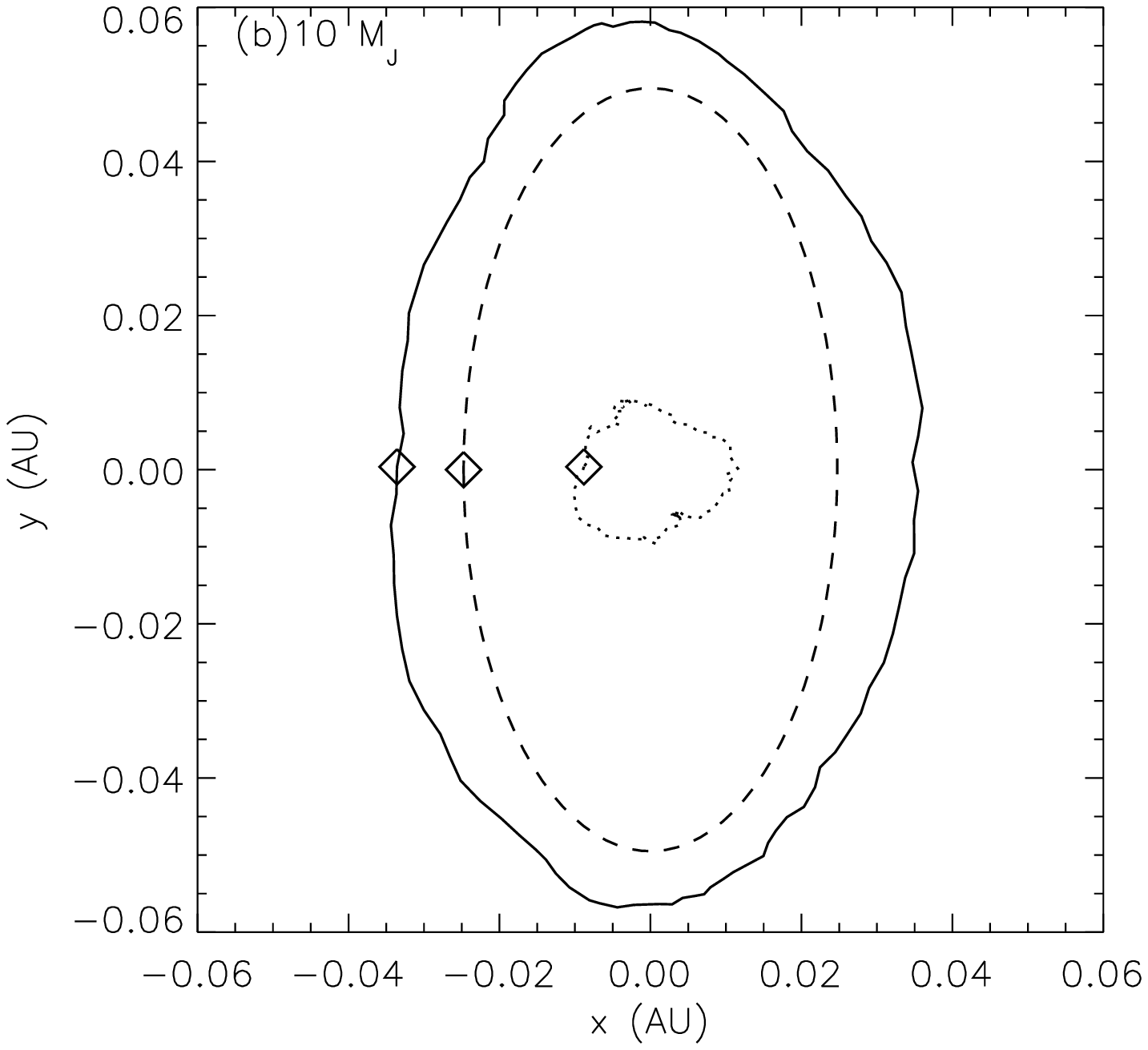}
\caption{
The shift of the photo-center with a low viscous disk $\alpha_{\rm
vis}=10^{-3}$. 
The disk is viewed at an inclined angle $v=60^{\circ}$.
The solid, dashed and dotted lines represent shifts in photo-centers
(see Fig. \ref{fig:cl_std})
($a$) The photo-center motions with a Jupiter mass planet ($M_p=M_J$).
($b$) The motions with a 10 Jupiter mass planet ($M_p=10M_J$).
\label{fig:cl_a1D-3}
}
\end{figure}

\begin{figure}
\epsscale{2.2}
\plottwo{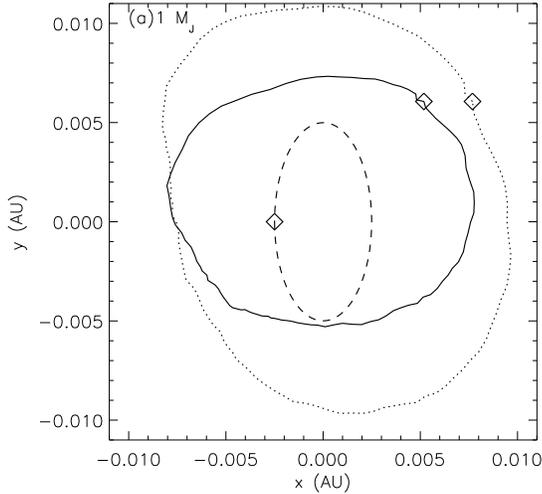}{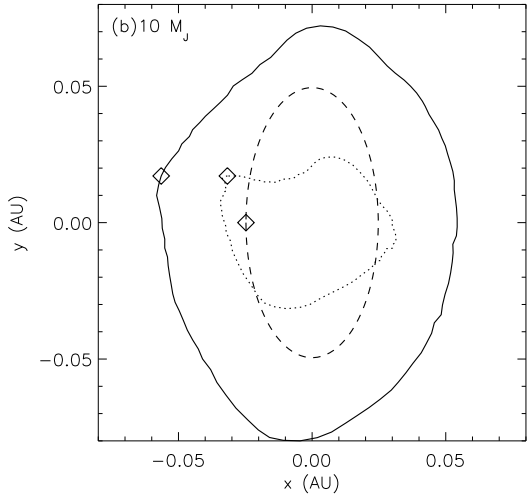}
\caption{
The shift of the photo-center with a narrow gap between $[0.8r_p, 1.2r_p]$.
The disk is viewed at an inclined angle $v=60^{\circ}$.
The solid, dashed and dotted lines represent shifts in photo-centers
(see Fig. \ref{fig:cl_std})
($a$) The photo-center motions with a Jupiter mass planet ($M_p=M_J$).
($b$) The motions with a 10 Jupiter mass planet ($M_p=10M_J$).
\label{fig:cl_ngap}
}
\end{figure}

\begin{figure}
\epsscale{1.0}
\plotone{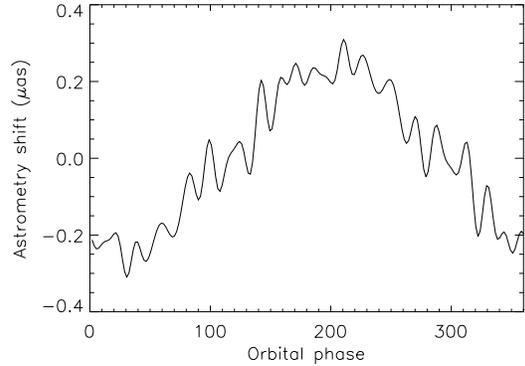}
\caption{
The predicted astronomy shift by the SIM is plotted against the orbital
phase of the planet.
The disk is 140 pc away and observed at face-on. A Jupiter mass
planet orbits at 5 AU from a solar mass star.
The planet's orbital phase is measured from the baseline position angle
of the SIM.
\label{fig:sim_i0}
}
\end{figure}

\begin{figure}
\epsscale{1.0}
\plotone{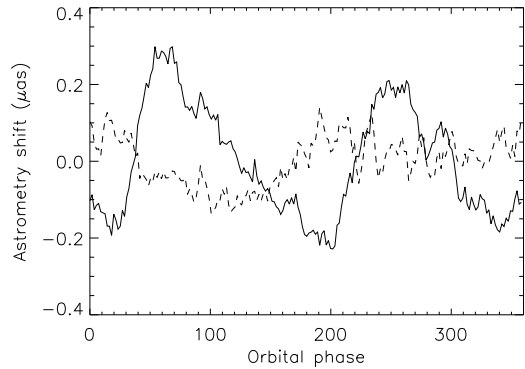}
\caption{
The predicted astronomy shift by the SIM is plotted against the orbital
phase of the planet.  The disk is observed at an inclination
$v=60^{\circ}$. 
The solid line shows the shift that is measured with a baseline parallel
to the minor axis of the disk image, while
the dashed line corresponds to a baseline parallel
to the major axis.
The planet's orbital phase is measured from the minor axis of the
disk image.
The disk is 140 pc away and a Jupiter mass
planet orbits at 5 AU from a solar mass star.
\label{fig:sim_i60}
}
\end{figure}

\begin{figure}
\epsscale{1.0}
\plotone{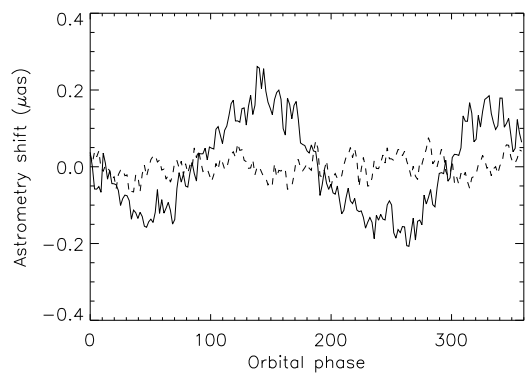}
\caption{
Same as Fig. \ref{fig:sim_i60}, but for
a Jupiter mass planet at 10 AU from a solar mass star.
\label{fig:sim_rp10_i60}
}
\end{figure}

\end{document}